\documentclass[aps,prx,superscriptaddress,twocolumn,letterpaper]{revtex4}

\usepackage[english]{babel}
\usepackage{xcolor,graphicx}

\usepackage{amsmath}
\usepackage{amssymb}
\usepackage{amsbsy}

\usepackage[colorlinks=true,linkcolor=blue,citecolor=blue,filecolor=blue,urlcolor=blue]{hyperref}

\DeclareMathOperator{\sgn}{sgn}
\DeclareMathOperator{\im}{Im}
\DeclareMathOperator{\re}{Re}

\newcommand{\beginsupplement}{%
		\clearpage
		\onecolumngrid
        \setcounter{table}{0}
        \renewcommand{\thetable}{S\arabic{table}}%
        \setcounter{figure}{0}
        \renewcommand{\thefigure}{S\arabic{figure}}%
        \setcounter{section}{0}
        \renewcommand{\thesection}{S\arabic{section}}%
        \setcounter{equation}{0}
        \renewcommand{\theequation}{S\arabic{equation}}%
     }

\begin{document}

\title{Light-mediated strong coupling between a mechanical oscillator \\and atomic spins one meter apart}

\author{Thomas M. Karg}
\affiliation{Department of Physics and Swiss Nanoscience Institute, University of Basel, Klingelbergstrasse 82, 4056 Basel, Switzerland}

\author{Baptiste Gouraud}
\altaffiliation{Present address: iXblue, 34 rue de la Croix de Fer, 78105 Saint-Germain-en-Laye, France}
\affiliation{Department of Physics and Swiss Nanoscience Institute, University of Basel, Klingelbergstrasse 82, 4056 Basel, Switzerland}

\author{Chun Tat Ngai}
\affiliation{Department of Physics and Swiss Nanoscience Institute, University of Basel, Klingelbergstrasse 82, 4056 Basel, Switzerland}

\author{Gian-Luca Schmid}
\affiliation{Department of Physics and Swiss Nanoscience Institute, University of Basel, Klingelbergstrasse 82, 4056 Basel, Switzerland}

\author{Klemens Hammerer}
\affiliation{Institute for Theoretical Physics and Institute for Gravitational Physics (Albert Einstein Institute), Leibniz Universit\"at Hannover, Appelstra\ss e 2, 30167 Hannover, Germany}

\author{Philipp Treutlein}
\email{philipp.treutlein@unibas.ch}
\affiliation{Department of Physics and Swiss Nanoscience Institute, University of Basel, Klingelbergstrasse 82, 4056 Basel, Switzerland}

\date{\today}

\begin{abstract}
Engineering strong interactions between quantum systems is essential for many phenomena of quantum physics and technology. Typically, strong coupling relies on short-range forces or on placing the systems in high-quality electromagnetic resonators, restricting the range of the coupling to small distances. 
We use a free-space laser beam to strongly couple a collective atomic spin and a micromechanical membrane over a distance of one meter in a room-temperature environment. 
The coupling is highly tunable and allows the observation of normal-mode splitting, coherent energy exchange oscillations, two-mode thermal noise squeezing and dissipative coupling. 
Our approach to engineer coherent long-distance interactions with light makes it possible to couple very different systems in a modular way, opening up a range of opportunities for quantum control and coherent feedback networks.
\end{abstract}


\maketitle

\let\oldaddcontentsline\addcontentsline
\renewcommand{\addcontentsline}[3]{}

Many of the recent breakthroughs in quantum science and technology rely on engineering strong, controllable interactions between quantum systems. In particular, Hamiltonian interactions that generate reversible, bidirectional coupling play an important role for creating and manipulating non-classical states in quantum metrology \cite{Pezze2018}, simulation \cite{Gross2017}, and information processing \cite{Ladd2010}.
For systems in close proximity, strong Hamiltonian coupling is routinely achieved, prominent examples being atom-photon coupling in cavity quantum electrodynamics \cite{Kimble2008} and coupling of trapped ions \cite{Blatt2008} or solid-state spins \cite{Hanson2008} via short-range electrostatic or magnetic forces. At macroscopic distances, however, the observation of strong Hamiltonian coupling is not only hampered by a severe drop in the interaction strength, but also by the fact that it becomes increasingly difficult to prevent information leakage from the systems to the environment, which renders the interaction dissipative \cite{Buchmann2015}. 
Overcoming these challenges would make Hamiltonian interactions available for reconfigurable long-distance coupling in quantum networks \cite{Kimble2008} and hybrid quantum systems \cite{Treutlein2014,Kurizki2015}, which so far employ mostly measurement-based or dissipative interactions. 

A promising strategy to reach this goal uses one-dimensional waveguides or free-space laser beams over which quantum systems can couple via the exchange of photons. 
Such cascaded quantum systems \cite{Gardiner2004} have attracted interest in the context of chiral quantum optics \cite{Lodahl2017,Chang2018} and waveguide quantum-electrodynamics \cite{Lalumiere2013}. 
A fundamental challenge in this approach is that the same photons that generate the coupling eventually leak out, thus allowing the systems to decohere at an equal rate. 
For this reason, light-mediated coupling is mainly seen today as a means for unidirectional state-transfer \cite{Ritter2012,Campagne-Ibarcq2018,Kurpiers2018}, or entanglement generation by collective measurement \cite{Julsgaard2001,Hofmann2012,Riedinger2018} or dissipation \cite{Krauter2011}. 
Decoherence by photon loss can be suppressed if the waveguide is terminated by mirrors to form a high quality resonator, which has enabled coherent coupling of superconducting qubits \cite{Majer2007,Mirhosseini2019}, atoms \cite{Baumann2010}, or atomic mechanical oscillators \cite{Spethmann2015} in mesoscopic setups. However, stability constraints and bandwidth limitations make it difficult to extend resonator-based approaches to larger distances. Strong bidirectional Hamiltonian coupling mediated by light over a truly macroscopic distance has so far remained elusive.

We pursue an alternative approach to realize long-distance Hamiltonian interactions which relies on connecting two systems by a laser beam in a loop geometry \cite{Kockum2018,Karg2019}. Through the loop the systems can exchange photons, realizing a bidirectional interaction. Moreover, the loop leads to an interference of quantum noise introduced by the light field. For any system that couples to the light twice and with opposite phase, quantum noise interferes destructively and associated decoherence is suppressed. At the same time information about that system is erased from the output field. In this way the coupled systems can effectively be closed to the environment, even though the light field mediates strong interactions between them.
Since the coupling is mediated by light, it allows systems of different physical nature to be connected over macroscopic distances. Furthermore, by manipulating the light field in between the systems, one can reconfigure the interaction without having to modify the quantum systems themselves. 
These features will be useful for quantum networking \cite{Kimble2008}.

We use this scheme to couple a collective atomic spin and a micromechanical membrane held in separate vacuum chambers, realizing a hybrid atom-optomechanical system \cite{Treutlein2014}. First experiments with such setups recently demonstrated sympathetic cooling \cite{Joeckel2015,Christoph2018}, quantum back-action evading measurement \cite{Moller2017} and entanglement \cite{Thomas2020}. Here, we realize strong Hamiltonian coupling and demonstrate the versatility of light-mediated interactions: we engineer beam-splitter and parametric-gain Hamiltonians and switch from Hamiltonian to dissipative coupling by applying a phase shift to the light field between the systems. This high level of control in a modular setup gives access to a unique toolbox for designing hybrid quantum systems \cite{Kurizki2015} and coherent feedback loops for advanced quantum control strategies \cite{Zhang2017}.

\section*{Description of the coupling scheme}

\begin{figure*}[tb!]
\includegraphics[width=0.66\linewidth]{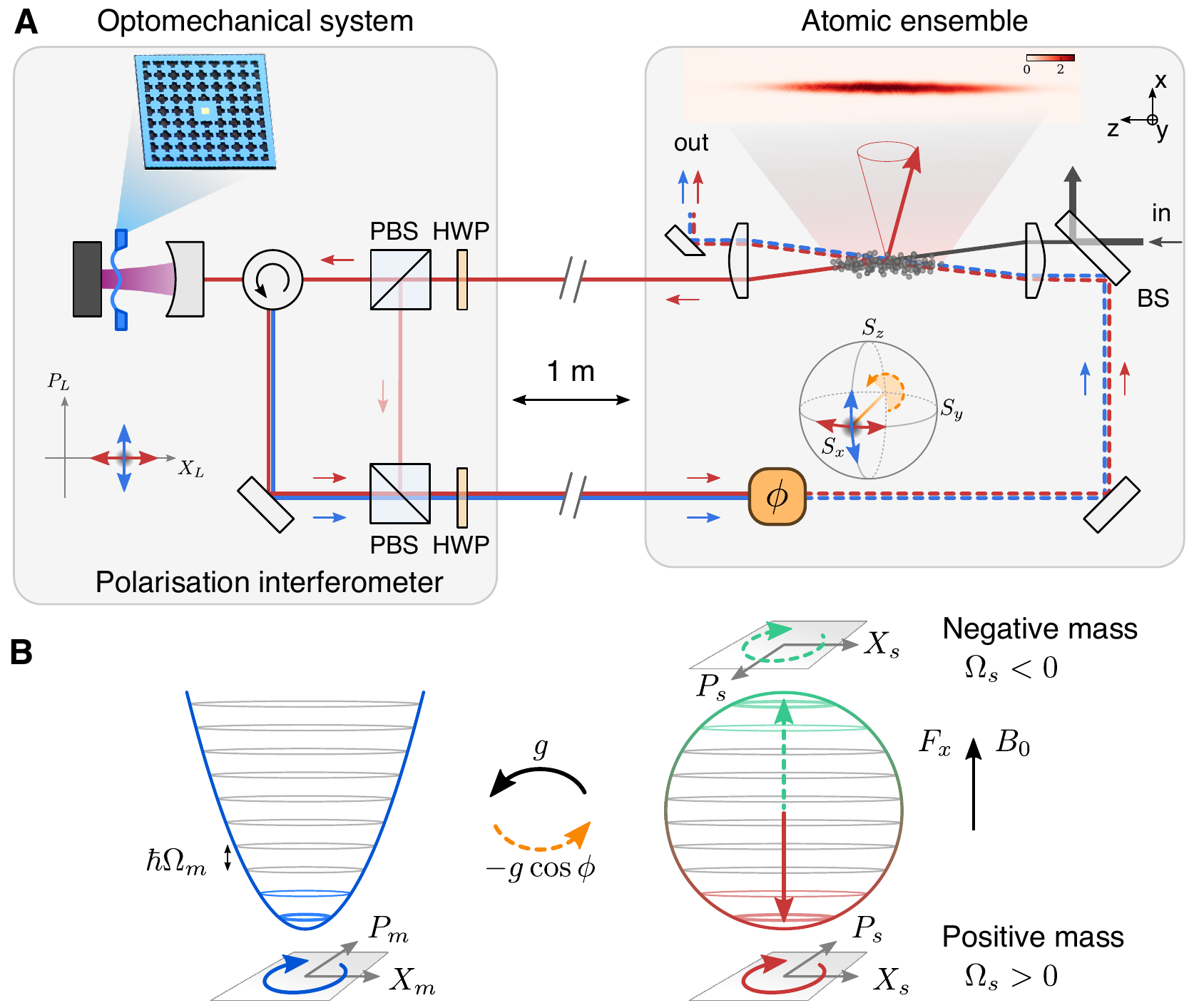}
\caption{Schematic setup for long-distance Hamiltonian coupling. (\textbf{A}), Cascaded coupling of an atomic spin ensemble (right) and a micromechanical membrane (left) by a free-space laser beam. The pictures show the silicon-nitride membrane embedded in a silicon chip with phononic crystal structure and a side-view absorption image of the atomic cloud (color bar: optical density). The laser beam first carries information from the atoms to the membrane and then loops back to the atoms such that it mediates a bidirectional interaction. A polarization interferometer (PBS: polarizing beam-splitter, HWP: half-wave plate) maps between the Stokes vector $\mathbf{S}$ defining the polarization state of light at the atoms and field quadratures $X_L, P_L$ relevant for the optomechanical interaction. The loop phase $\phi$ is controlled by a rotation of $\mathbf{S}$ by an angle $\phi$ in the optical path from the membrane to the atoms. (\textbf{B}), Effective interaction. The membrane vibration mode (harmonic oscillator) is coupled to the collective spin of the atoms (represented on a sphere). If the mean spin is oriented along an external magnetic field $B_0$ to either the south or north pole of the sphere, its small-amplitude dynamics can be mapped onto a harmonic oscillator with positive or negative mass, respectively. The relative phase of the spin-to-membrane coupling constant $g$ and the membrane-to-spin coupling constant $-g\cos\phi$ defines whether the effective dynamics are Hamiltonian ($\phi = \pi$) or dissipative ($\phi = 0$).}
\label{fig_1}
\end{figure*}

In the experimental setup (Fig.~\ref{fig_1}A and Supplementary Materials (SM) section~\ref{sec_suppl_1}), the atomic ensemble consists of $N = 10^7$ laser-cooled Rubidium-87 atoms in an optical dipole trap. The atoms form a collective spin $\mathbf{F} = \sum_{i = 1}^{N} \mathbf{f}^{(i)}$ with $\mathbf{f}^{(i)}$ being the $f=2$ ground state spin vector of atom $i$. Optical pumping polarises $\mathbf{F}$ along an external magnetic field $\mathbf{B_0}$ in the $x$-direction such that the spin acquires a macroscopic orientation $\bar{F}_x = - f N$. The small-amplitude dynamics of the transverse spin components $F_y, F_z$ are well approximated by a harmonic oscillator \cite{Hammerer2010} with position $X_s = F_z / \sqrt{|\bar{F}_x|}$ and momentum $P_s = F_y /\sqrt{|\bar{F}_x|}$. It oscillates at the Larmor frequency $\Omega_s\propto B_0$, which is tuned by the magnetic field strength.
A feature of the spin system is that it can realize such an oscillator with either positive or negative effective mass \cite{Polzik2015,Moller2017}. This is achieved by reversing the orientation of $\mathbf{F}$ with respect to $\mathbf{B_0}$, which reverses the sense of rotation of the oscillator in the $X_s,P_s$-plane (Fig.~\ref{fig_1}B). This feature allows us to realize different Hamiltonian dynamics with the spin coupled to the membrane.

The spin interacts with the coupling laser beam through an off-resonant Faraday interaction \cite{Hammerer2010} $H_s = 2\hbar \sqrt{\Gamma_s/\bar{S}_x} X_s S_z$, which couples $X_s$ to the polarization state of the light, described by the Stokes vector $\mathbf{S}$. Initially, the laser is linearly polarized along $x$ with $\bar{S}_x = \Phi_L /2$, where $\Phi_L$ is the photon flux. The strength of the atom-light coupling depends on the spin measurement rate $\Gamma_s \propto d_0 \Phi_L/\Delta_a^2$, which is proportional to the optical depth $d_0\approx 300$ of the atomic ensemble (SM). Choosing a large laser-atom detuning $\Delta_a = -2\pi\times 80$~GHz suppresses spontaneous photon scattering while maintaining a sizable coupling. 

The mechanical oscillator is the $(2,2)$ square drum mode of a silicon-nitride membrane at a vibrational frequency of $\Omega_m = 2\pi\times1.957~$MHz with a quality factor of $1.3\times 10^6$ \cite{Thompson2008}. It is placed in a short single-sided optical cavity to enhance the optomechanical interaction while maintaining a large cavity bandwidth for fast and efficient coupling to the external light field. Radiation pressure couples the membrane displacement $X_m$ to the amplitude fluctuations $X_L$ of the light entering the cavity on resonance, with Hamiltonian $H_{m} = 2\hbar \sqrt{\Gamma_m} X_m X_L$ \cite{Aspelmeyer2014}. Here, we defined the optomechanical measurement rate $\Gamma_m = (4g_0/\kappa)^2 \Phi_m$ that depends on the vacuum optomechanical coupling constant $g_0$, cavity linewidth $\kappa$, and photon flux $\Phi_m$ entering the cavity (SM). In the present setup, the optomechanical cavity is mounted in a room temperature vacuum chamber, making thermal noise the dominant noise source of the experiment. 

The light-field mediates a bidirectional coupling between spin and membrane. A spin displacement $X_s$ is mapped by $H_s$ to a polarization rotation $S_y = 2\sqrt{\Gamma_s\bar{S}_x} X_s$ of the light. A polarization interferometer (Fig.~\ref{fig_1}A) converts this to an amplitude modulation $X_L \approx S_y/\sqrt{\bar{S}_x}$ at the optomechanical cavity, resulting in a force $\dot{P}_m = - 4\sqrt{\Gamma_m \Gamma_s} X_s$ on the membrane. Conversely, a membrane displacement $X_m$ is turned by $H_m$ into a phase-modulation $P_L = -2\sqrt{\Gamma_m}X_m$ of the cavity output field. The interferometer converts this to a polarization rotation $S_z \approx \sqrt{\bar{S}_x} P_L$, resulting in a force $\dot{P}_s = 4\sqrt{\Gamma_s\Gamma_m} X_m$ on the spin. A small angle between the laser beams in the two atom-light interactions prevents light from going once more to the membrane. Consequently, the cascaded setup promotes a bidirectional spin-membrane coupling. A fully quantum mechanical treatment (SM) confirms this picture and predicts a spin-membrane coupling strength $g = (\eta^2 + \eta^4)\sqrt{\Gamma_s \Gamma_m}$, accounting for an effective optical power transmission $\eta^2 \approx 0.8$ between the systems.

The light-mediated interaction can be thought of as a feedback loop that transmits a spin excitation to the membrane, whose response then acts back on the spin, and vice versa (Fig.~\ref{fig_1}B). After one round-trip, the initial signal has acquired a phase $\phi$, the loop phase. The discussion above refers to a vanishing loop phase $\phi = 0$ and shows that the forces $\dot{P}_m = -2gX_s$ and $\dot{P}_s = +2g X_m$ differ in their relative sign. Such a coupling is non-conservative and cannot arise from a Hamiltonian interaction. With full access to the laser beams, we can tune the loop phase by inserting a half-wave plate (HWP) in the path from the membrane back to the atoms, which rotates the Stokes vector by an angle $\phi = \pi$ about $S_x$. This inverts both $S_y$ and $S_z$, which carry the spin and membrane signals respectively, thus switching the dynamics to a fully Hamiltonian force, $\dot{P}_m = -2gX_s$ and $\dot{P}_s = -2gX_m$.

All these phenomena are unified in a rigorous quantum-mechanical theory \cite{Karg2019} of the cascaded light-mediated coupling, which also correctly describes the dynamics for an arbitrary loop phase. It allows us to describe the effective dynamics of the coupled spin-membrane system with density operator $\rho$ by a Markovian master equation
\begin{equation}
\dot{\rho} = \frac{1}{i\hbar} [H_0 + H_\mathrm{eff}, \rho]  - \frac{1}{2}\left(J^\dag J \rho + \rho J^\dag J\right) + J \rho J^\dag.
\end{equation}
Here, we neglect optical loss and light propagation delay between the systems for brevity. The dynamics consist of a unitary part with free harmonic oscillator Hamiltonian $H_0 = \sum_{i = s,m} \hbar \Omega_i (X_i^2 + P_i^2)/2$ and effective interaction Hamiltonian $H_\mathrm{eff} = (1-\cos\phi)\hbar g X_s X_m + 2 \sin(\phi) \hbar \Gamma_s X_s^2$, and a dissipative part with collective jump operator $J =  \sqrt{2\Gamma_m} X_m + i (1 + e^{i\phi})\sqrt{2\Gamma_s} X_s$. Next to the coherent spin-membrane coupling, $H_\mathrm{eff}$ also includes a spin self-interaction which vanishes for the specific cases $\phi = 0,\pi$ considered here. The jump operator contains a constant membrane term and a spin term that is modulated by $\phi$ due to interference of the two spin-light interactions. From the dependence of $H_\mathrm{eff}$ and $J$ on $\phi$, it is clear that $\phi = 0$ corresponds to vanishing Hamiltonian coupling and maximum dissipative coupling. Accordingly, we refer to $\phi = 0$ as the dissipative regime. On the other hand, $\phi = \pi$ maximizes the coherent spin-membrane coupling in $H_\mathrm{eff}$ and at the same time leads to destructive interference of the spin term in $J$, we thus call $\phi = \pi$ the Hamiltonian regime. 
Both regimes will be experimentally explored in the following, each with the atomic spin realizing either a positive- or negative-mass oscillator. This gives rise to a whole range of different dynamics in a single system, which can be harnessed for different purposes in quantum technology.

\section*{Results}

\subsection*{Normal-mode splitting}

We first investigate the light-mediated coupling in the Hamiltonian regime ($\phi = \pi$) and with the spin realizing a positive-mass oscillator. At a magnetic field of $B_0 = 2.81$~G the spin is tuned into resonance with the membrane ($\Omega_s = \Omega_m$). In this configuration, the resonant terms in $H_\mathrm{eff}$ realize a beam-splitter interaction $H_\mathrm{BS} = \hbar g (b_s^\dag b_m + b_m^\dag b_s)$, which generates state swaps between the two systems. Here $b_s = (X_s + i P_s) /\sqrt{2}$ and $b_m = (X_m + i P_m)/\sqrt{2}$ are annihilation operators of the spin and mechanical modes, respectively. 

\begin{figure}[tb!]
\includegraphics[width=\linewidth]{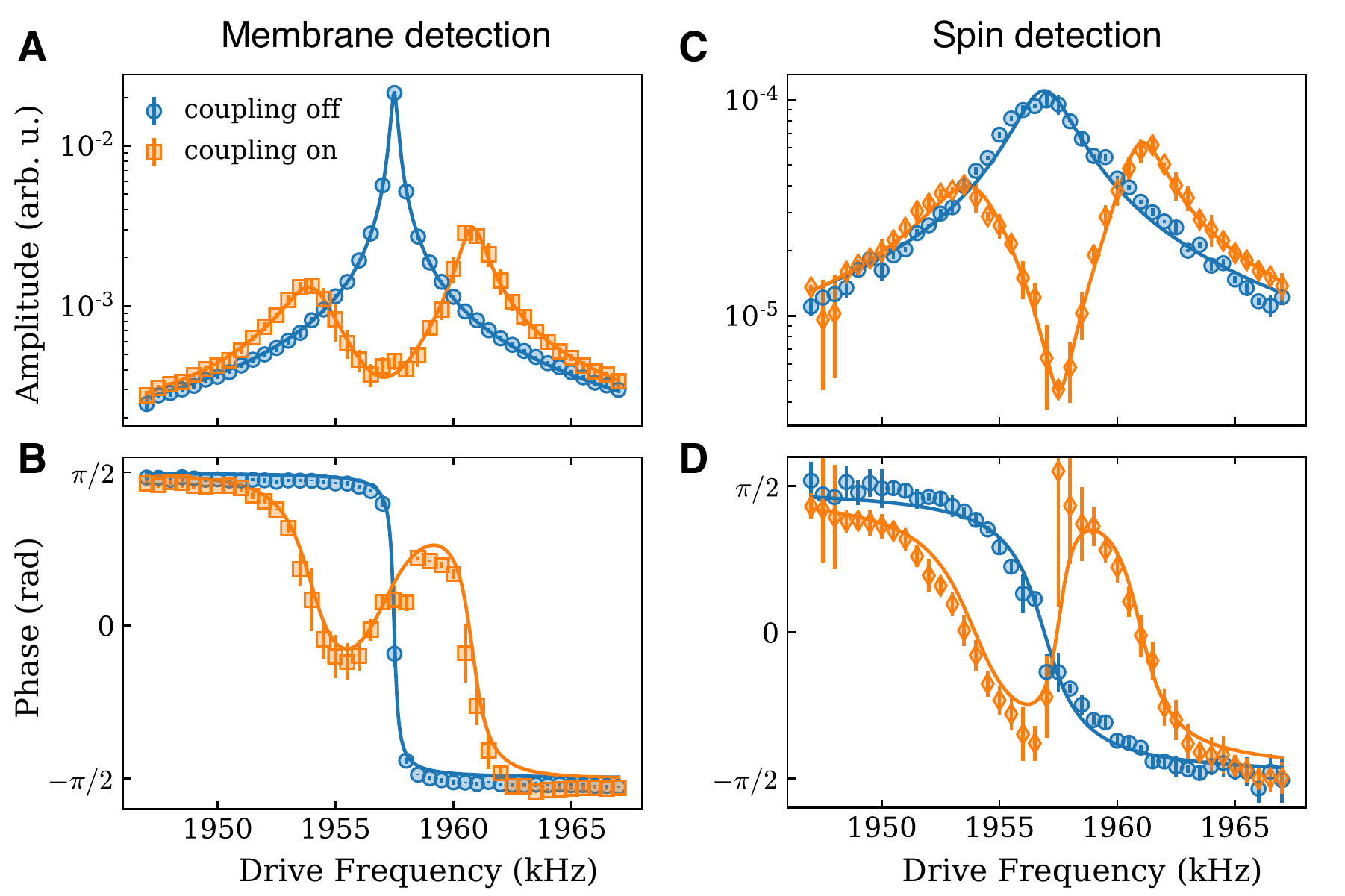}\\
\caption{Observation of strong spin-membrane coupling. Spectroscopy of the membrane (\textbf{A,B}) and the spin (\textbf{C,D}), both revealing a normal mode splitting if the coupling beam is on and the oscillators are resonant ($\Omega_s = \Omega_m$). For comparison we show the uncoupled responses of the membrane with coupling beam off ($\textbf{A,B}$) and of the spin with cavity off-resonant ($\textbf{C,D}$). Lines are fits to the data with a coupled-mode model (SM). Error bars are standard deviations of 3 independent measurements.}
\label{fig_2}
\end{figure}

We perform spectroscopy of the coupled system using independent drive and detection channels for spin and membrane. The membrane vibrations are recorded by balanced homodyne detection using an auxiliary laser beam coupled to the cavity in orthogonal polarization. To drive the membrane, this beam is amplitude modulated. 
The spin precession is detected by splitting off a small portion of the coupling light on the path from spin to membrane. 
A radio-frequency (RF) magnetic coil drives the spin.
We measure the amplitude and phase response of either system using a lock-in amplifier that demodulates the detector signal at the drive frequency (see SM section~\ref{sec_suppl_2}). After spin-state initialization we simultaneously switch on coupling and drive and start recording. The drive frequency is kept fixed during each experimental run and stepped between consecutive runs. 

Figs.~\ref{fig_2}A and B show the membrane's response in amplitude and phase, respectively. With the coupling beam off, it exhibits a Lorentzian resonance of linewidth $\gamma_m = 2\pi\times0.3$~kHz, broader than the intrinsic linewidth due to optomechanical damping by the red-detuned cavity field \cite{Aspelmeyer2014}. For the uncoupled spin oscillator (Figs.~\ref{fig_2}C, D) with cavity off-resonant, we also measure a Lorentzian response of linewidth $\gamma_s = 2\pi \times 4$~kHz, broadened by the coupling light.
When we turn on the coupling to the spin, the membrane resonance splits into two hybrid spin-mechanical normal modes. This signals strong coupling \cite{Groeblacher2009,Verhagen2012}, where light-mediated coupling dominates over local damping. 
Fitting the well-resolved splitting yields $2g = 2\pi \times 6.1$~kHz, which exceeds the average linewidth $(\gamma_s + \gamma_m)/2 = 2\pi\times 2$~kHz and agrees with the expectation based on an independent calibration of the systems (SM). 
A characteristic feature of the long-distance coupling is a finite delay $\tau$ between the systems. It causes a linewidth asymmetry of the two normal modes when $\Omega_s = \Omega_m$, which we observe in Fig.~\ref{fig_2}. The fits yield a value of $\tau = 15$~ns, consistent with the propagation delay of the light between the systems and the cavity response time.

We also observe normal-mode splitting in measurements of the spin (Figs.~\ref{fig_2}C and D). Here, the combination of the broader spin linewidth with the much narrower membrane resonance results in a larger dip between the two normal modes and a larger phase shift, in analogy to optomechanically-induced transparency \cite{Weis2010}.

\subsection*{Energy exchange oscillations}

\begin{figure}[tb!]
\includegraphics[width=\linewidth]{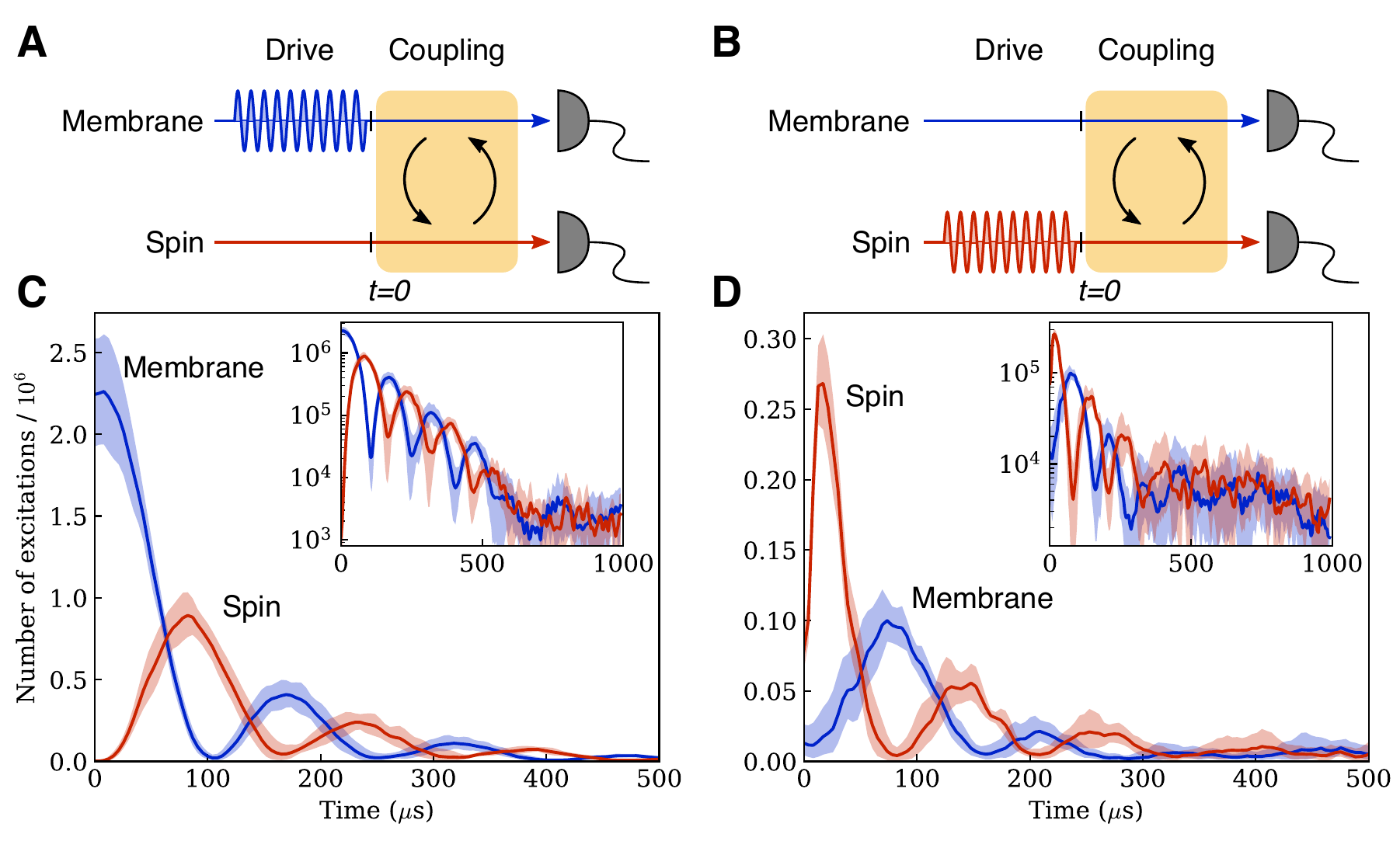}\\
\caption{Time-domain exchange oscillations showing coherent energy transfer between spin and membrane. (\textbf{A}), Pulse sequence for excitation of the membrane by radiation-pressure modulation via the auxiliary laser beam. (\textbf{B}), Pulse sequence for spin excitation with an external RF magnetic field. (\textbf{C}), Oscillations in the excitation numbers of membrane and spin as a function of the interaction time, measured using the pulse sequence in A. (\textbf{D}), Data obtained with pulse sequence B and weaker drive strength than in C. Here, the finite rise time of the spin signal at $t=0$ corresponds to the turn-on of the coupling beam, which is also used for spin detection. Insets in C,D show the same data on a log-scale. Lines and shaded areas represent the mean and one standard deviation of five measurements, respectively.}
\label{fig_3}
\end{figure}

Having observed the spectroscopic signature of strong coupling, we now use it for swapping spin and mechanical excitations in a pulsed experiment. We start by coherently exciting the membrane to $\approx 2\times10^6$ phonons, a factor of $100$ above its mean equilibrium energy, by applying an amplitude modulation pulse to the auxiliary cavity beam (Fig.~\ref{fig_3}A). At the same time, the spin is prepared in its ground state with $\Omega_s = \Omega_m$. 
The coupling beam is switched on at time $t=0~\mu$s and the displacements $X_{s}(t)$ and $X_m(t)$ of spin and membrane are continuously monitored via the independent detection. From the measured mean square displacements we determine the excitation number of each system (SM).
Fig.~\ref{fig_3}C shows the excitation numbers as a function of the interaction time. The data show coherent and reversible energy exchange oscillations from the membrane to the spin and back with an oscillation period of $T \approx 150~\mu$s, in accordance with the value $\pi/g$ extracted from the observed normal-mode splitting. Damping limits the maximum energy transfer efficiency at time $T/2$ to about 40\%. 

The same experiment is repeated but with the initial drive pulse applied to the spin (Figs.~\ref{fig_3}B and D). Here, we observe another set of exchange oscillations with the same periodicity, swapping an initial spin excitation of $n_s \approx 3\times 10^5$ to the membrane and back. After the coherent dynamics have decayed, the systems equilibrate in a thermal state of $\approx 3\times 10^3$ phonons, lower than the effective optomechanical bath of $1.5\times10^4$ phonons, demonstrating sympathetic cooling \cite{Joeckel2015} of the membrane by the spin. The observed sympathetic cooling strength agrees with simulations using the experimentally determined parameters.

\subsection*{Parametric-gain dynamics}

So far we have explored Hamiltonian coupling of the membrane to a spin oscillator with positive effective mass, where the resonant interaction is of the beam-splitter type. 
If instead we reverse the magnetic field to $B_0 = -2.81$~G but keep the spin pumping direction the same, the collective spin is prepared in its highest energy state with $\bar{F}_x = +N f$. In this case any excitation reduces the energy such that the spin oscillator has a negative effective mass \cite{Julsgaard2001} and $\Omega_s = -\Omega_m$ (Fig.~\ref{fig_1}B). The resonant term of $H_\mathrm{eff}$ is now the parametric-gain interaction \cite{Clerk2010} $H_\mathrm{PG} = \hbar g (b_s b_m + b_s^\dag b_m^\dag)$, which generates correlations between the two systems. 

\begin{figure}[tb!]
\includegraphics[width=\linewidth]{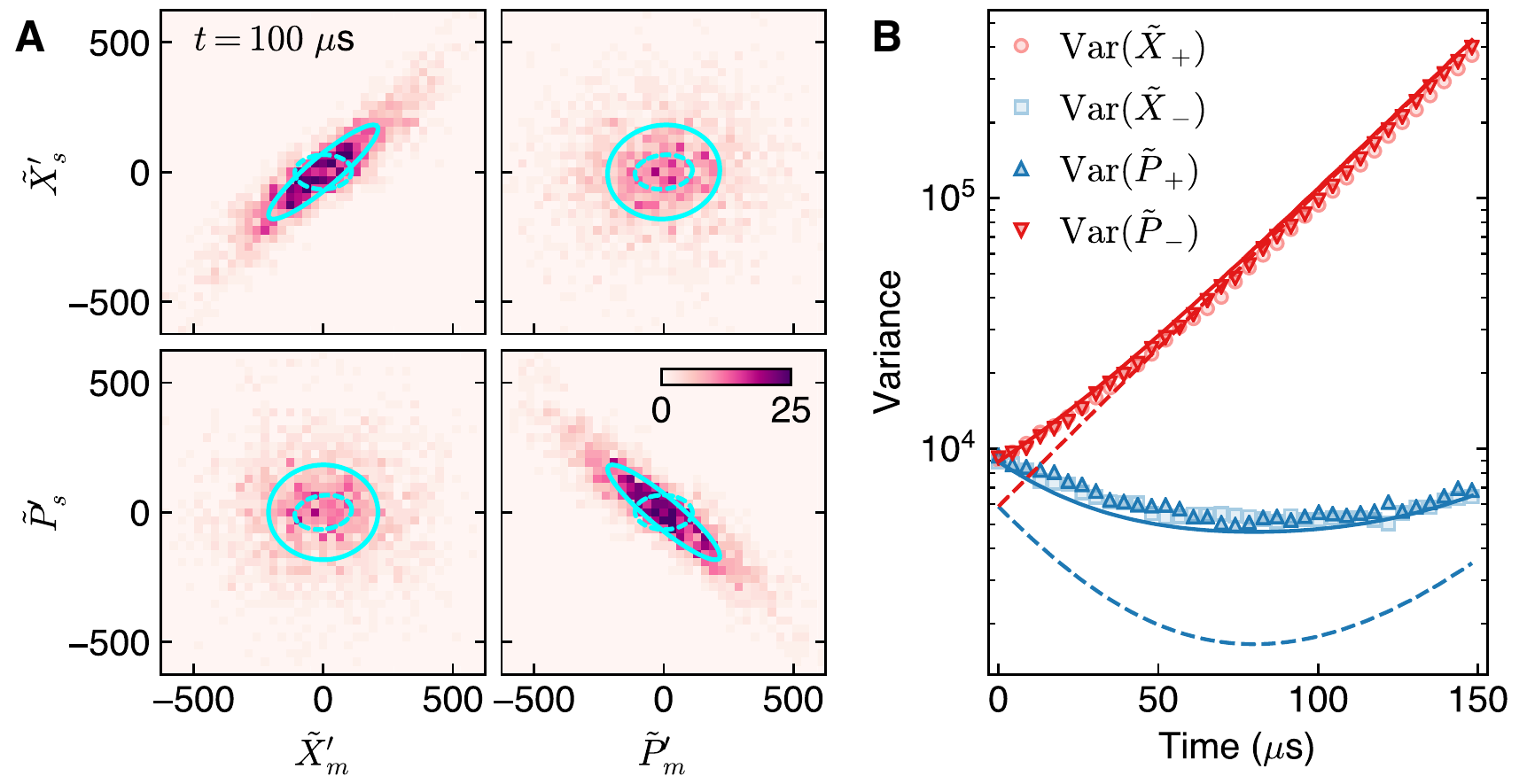}
\caption{Dynamics of the parametric-gain interaction with thermal noise averaged over 2000 realizations. (\textbf{A}), Phase space histograms showing correlations between the rotated spin and membrane quadratures after $100~\mu$s of interaction time. The solid (dashed) ellipses enclose regions of one standard deviation at $t=100~\mu$s ($t=0~\mu$s). (\textbf{B}), Variances of the combined quadratures $\tilde{X}_{\pm}$ and $\tilde{P}_{\pm}$ as a function of interaction time. Exponential increase is observed for quadratures $\tilde{X}_{+}$ and $\tilde{P}_{-}$ while noise reduction is measured for $\tilde{X}_{-}$ and $\tilde{P}_{+}$. The solid lines are a simulation of the corresponding variances including a spin detector noise floor of $6\times 10^3$, while the dashed lines assume noise-free detection.}
\label{fig_4}
\end{figure}

We investigate the dynamics generated by $H_\mathrm{PG}$ with the membrane driven by thermal noise. In order to quantify the development of spin-mechanical correlations, we determine slowly varying quadratures $\tilde{X}_{s,m}'$ and $\tilde{P}_{s,m}'$ of both systems as the cosine and sine components of the demodulated detector signals, respectively (SM). Adjusting the demodulator phase allows us to find the basis with strongest correlations.
Fig.~\ref{fig_4}A shows histograms of the measured spin-mechanical correlations after an interaction time of $t=100~\mu$s. In each subplot, the dashed ellipse corresponds to the Gaussian 1-sigma contour of the measured histogram at $t= 0~\mu$s while the solid line is the contour at $t=100~\mu$s. Compared to the uncorrelated initial state, the histograms show strong amplification along the axes $\tilde{X}_+ = (\tilde{X}_s' + \tilde{X}_m')/\sqrt{2}$ and $\tilde{P}_- = (\tilde{P}_s' - \tilde{P}_m')/\sqrt{2}$, and a small amount of thermal noise squeezing along $\tilde{X}_- = (\tilde{X}_s' - \tilde{X}_m')/\sqrt{2}$ and $\tilde{P}_+ = (\tilde{P}_s' + \tilde{P}_m')/\sqrt{2}$. The quadrature pairs $\tilde{X}_{s}',\tilde{P}_{m}'$ and $\tilde{P}_{s}',\tilde{X}_{m}'$ remain uncorrelated.

In the time evolution of the combined variances $\tilde{X}_\pm$ and $\tilde{P}_\pm$ (Fig.~\ref{fig_4}B), at $t=0$ all variances start from the same value indicating an uncorrelated state. As time evolves, the variances of $\tilde{X}_{+}$ and $\tilde{P}_{-}$ grow exponentially, demonstrating the dynamical instability in this configuration, while $\tilde{X}_{-}$ and $\tilde{P}_{+}$ are squeezed and reach a minimum at $t=80~\mu$s before they grow again. The exponential growth rate of $2\pi \times 4.5~$kHz is consistent with the value of $2g - (\gamma_m + \gamma_s)/2$ extracted from the normal-mode splitting. 
For comparison, we also show simulated variances for the experimental parameters which are given by the lines in Fig.~\ref{fig_4}B (SM). Good agreement between data and simulation is found when accounting for a spin detector noise floor of $6\times 10^3$ (solid lines). The dashed lines correspond to perfect detection and show thermal noise squeezing by 5.5~dB.
Realizing the parametric-gain interaction by light-mediated coupling represents an important step towards generation of spin-mechanical entanglement by two-mode squeezing across macroscopic distances. Such entanglement is useful for metrology beyond the standard quantum limit \cite{Pezze2018}.

\subsection*{Control of the loop phase}

\begin{figure*}[t!]
\includegraphics[width=0.8\linewidth]{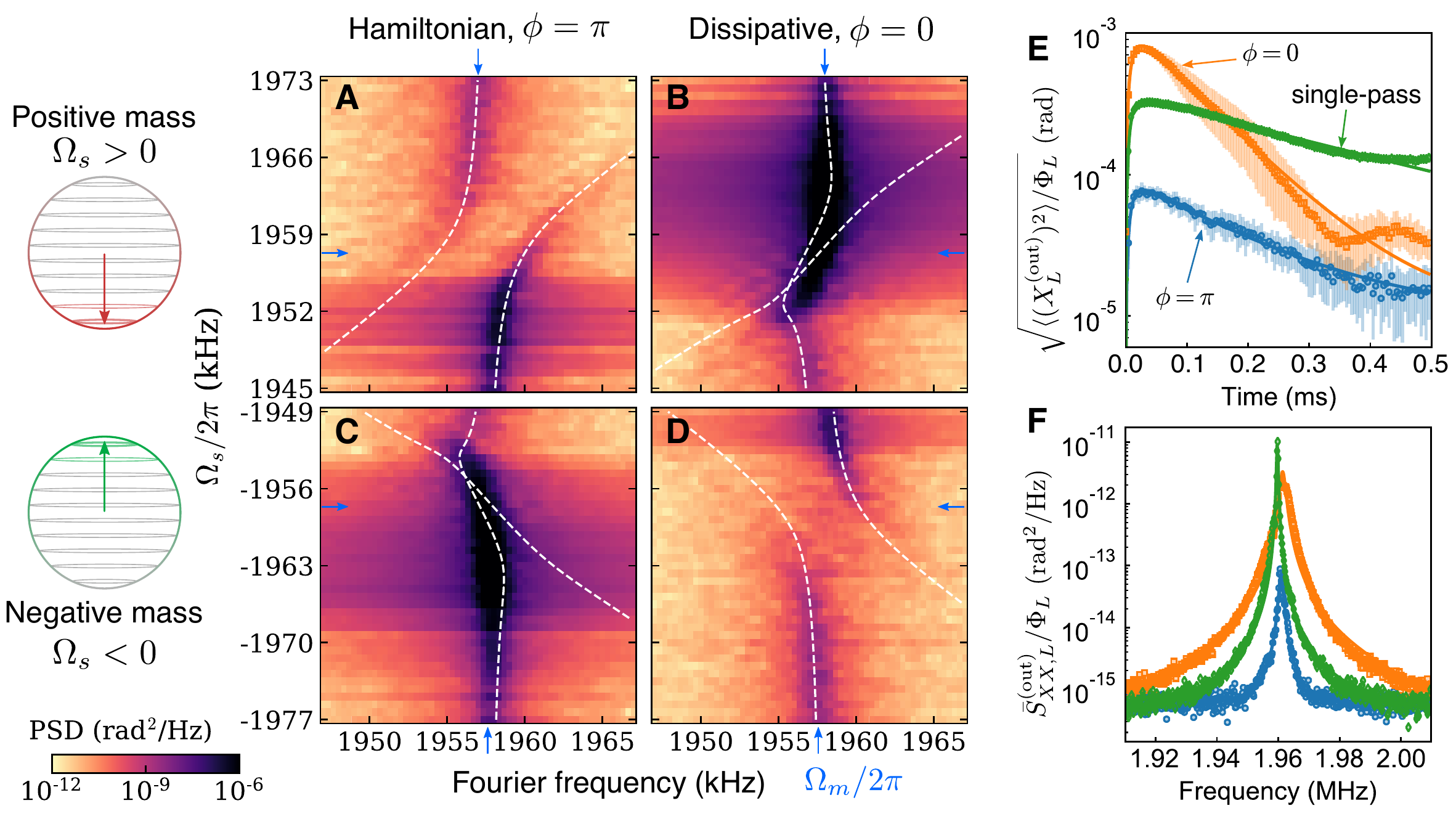}
\caption{Control of the loop phase. (\textbf{A--D}) Density plots of the membrane's thermal noise spectra in four different regimes, with membrane Fourier frequency on the horizontal axis ($\Omega_m$ indicated by blue arrows) and spin frequency $\Omega_s$ (controlled by magnetic field) on the vertical axis. Dashed white lines are the calculated normal mode frequencies. (\textbf{A}), Hamiltonian coupling with positive-mass spin oscillator (beam-splitter interaction): an avoided crossing is observed. (\textbf{B}), Dissipative coupling with positive-mass spin oscillator: level attraction and unstable dynamics at the exceptional point. (\textbf{C}), Hamiltonian coupling with negative-mass spin oscillator (parametric-gain interaction): unstable dynamics and an exceptional point. (\textbf{D}), Dissipative coupling with negative-mass spin oscillator: observation of an avoided crossing. (\textbf{E}), atomic spin signal (RMS amplitude) on the output light after pulsed excitation: constructive (destructive) interference of the two atom-light interactions is observed for $\phi = 0$ ($\phi = \pi$) compared to a single-pass interaction. The membrane is decoupled by detuning the cavity. Error bars are standard deviations of 25 repetitions. (\textbf{D}), frequency-domain power spectra corresponding to the data of \textbf{E}.}
\label{fig_5}
\end{figure*}

Equipped with control over both the loop phase and the effective mass of the spin oscillator, we can access four different regimes of the spin-membrane coupling: two Hamiltonian configurations with $\phi = \pi$ and $\Omega_s = \pm \Omega_m$, and the two corresponding dissipative configurations where we set $\phi = 0$ by omitting the half-wave plate in the optical path from membrane to atoms (SM).
While the dynamics in these configurations are fundamentally different and have different quantum noise properties, we obtain simple equations of motion for the expectation values,
\begin{eqnarray}
\ddot{X}_m + \gamma_m \dot{X}_m + \Omega_m^2 X_m &=& - g \Omega_m X_s(t-\tau),\\
\ddot{X}_s + \gamma_s \dot{X}_s +\Omega_s^2 X_s  &=& + g \Omega_s \cos(\phi) X_m(t-\tau),
\end{eqnarray}
with the damped harmonic oscillations on the left and the delayed coupling terms on the right. These are derived from Heisenberg-Langevin equations of the full system (see SM section~\ref{sec_suppl_4}) and reproduce the dynamics of the master equation in the limit $\tau\to0$. Two distinct regimes can be identified. If $\Omega_s \cos\phi < 0$ we expect stable dynamics equivalent to a beam-splitter interaction. In the opposite case where $\Omega_s \cos\phi > 0$, the dynamics are equivalent to a parametric-gain interaction and unstable. A simultaneous sign reversal of $\Omega_s$ and a $\pi$-shift of $\phi$ should leave the dynamics invariant.

To probe the dynamics in these configurations, we record thermal noise spectra of the membrane while the spin Larmor frequency is tuned across the mechanical resonance $\Omega_m = 2\pi\times 1.957$ MHz.
The Hamiltonian configuration with positive-mass spin oscillator is depicted in Fig.~\ref{fig_5}A, showing an avoided crossing at $\Omega_s = \Omega_m$ with frequency splitting $2g = 2\pi \times 5.9$~kHz, as in Fig.~\ref{fig_2} above. 
The dashed lines are the calculated normal mode frequencies (SM). 
The enhancement of the mechanical noise power for $\Omega_s < \Omega_m$ as compared to increased damping for $\Omega_s > \Omega_m$ is again a consequence of the finite optical propagation delay $\tau$ modifying the damping (SM).

Switching to the dissipative regime with $\phi = 0$ renders the system unstable due to positive feedback of the coupled oscillations (Fig.~\ref{fig_5}B). Instead of an avoided crossing, the normal modes are now attracted and cross near $\Omega_s = 2\pi\times 1.953$~MHz, forming one strongly amplified and one strongly damped mode. The former leads to exponential growth of correlated spin-mechanical motion, finally resulting in limit-cycle oscillations which dominate the power spectrum. This ensues a breakdown of the coupled oscillator model, such that the observed spectral peak shifts towards the unperturbed mechanical resonance. Still, the data are in good agreement with the theoretical model.

In Fig.~\ref{fig_5}C,D we repeat the experiments of Fig.~\ref{fig_5}A,B with negative-mass spin oscillator. The data show that Hamiltonian coupling with negative-mass spin oscillator produces similar spectra as dissipative coupling with positive-mass spin oscillator. In these configurations, the coupled system features an exceptional point \cite{Xu2016} where the normal modes become degenerate \cite{Bernier2018} and define the squeezed and anti-squeezed quadratures. Conversely, dissipative coupling together with an inverted spin (D) shows an avoided crossing with similar parameters as in the Hamiltonian case (A). This equivalence at the level of the expectation values is expected to break down once quantum noise of the light becomes relevant. Due to interference in the loop, quantum back-action on the spin is suppressed in the Hamiltonian coupling configuration, but enhanced in the dissipative configuration. 

A necessary condition for quantum back-action cancellation is destructive interference of the spin signal in the output field (see SM section~\ref{sec_suppl_3}). Fig.~\ref{fig_5}E and D show homodyne measurements of coherent spin precession on the coupling beam output quadrature $X_L^\mathrm{(out)}$ in time and frequency-domain, respectively. Toggling the loop phase between $\phi = 0$ and $\phi = \pi$, we observe a large interference contrast $> 10$ in the root-mean-squared (RMS) spin signal, showing that a spin measurement made by  light in the first pass can be erased in the second pass. Optical loss of $1-\eta^4 \approx 0.35$ inside the loop allows some information to leak out to the environment and brings in uncorrelated noise, limiting the achievable back-action suppression. Full interference in the output is still observed because the carrier and signal fields are subject to the same losses.
Since this principle of quantum back-action interference is fully general, it could be harnessed as well for other optical or microwave photonic networks \cite{Kimble2008,Kockum2018}.

\section*{Conclusion}

The observed normal-mode splitting and coherent energy exchange oscillations establish strong spin-membrane coupling, where the coupling strength $g$ exceeds the damping rates of both systems \cite{Groeblacher2009}. In order to achieve quantum-coherent coupling \cite{Verhagen2012}, $g$ must also exceed all thermal and quantum back-action decoherence rates. This will make it possible to swap non-classical states between the systems or to generate remote entanglement by two-mode squeezing. 
Thermal noise on the mechanical oscillator is the major source of decoherence in our room-temperature setup. We expect that modest cryogenic cooling of the optomechanical system to 4~K together with an improved mechanical quality factor of $>10^7$ \cite{Tsaturyan2017} will enable quantum-limited operation (see SM section~\ref{sec_suppl_4}). The built-in suppression of quantum back-action in the Hamiltonian configuration is a crucial feature of our coupling scheme. Interference of the two spin-light interactions reduces the spin's quantum back-action rate to $\gamma_{s,\mathrm{ba}} = (1-\eta^4)\Gamma_s$ while it is $\gamma_{m,\mathrm{ba}} = \eta^2 \Gamma_m$ for the membrane. Assuming thermal noise is negligible, the quantum cooperativity $C = 2g / (\gamma_{s,\mathrm{ba}} + \gamma_{m,\mathrm{ba}})$ can be optimized for a given one-way transmission $\eta^2$. We find an upper bound $C \leq \eta(1 + \eta^2)/\sqrt{1-\eta^4}$, reaching $2.7$ for our current setup. The bound is achieved for an optimal choice of measurement rates $\Gamma_s/\Gamma_m = \eta^2/(1-\eta^4)$, balancing the back-action on both systems. Further improvement is possible with a double-loop coupling scheme that also suppresses quantum back-action on the membrane \cite{Karg2019}. In this case, $C = \eta / (1-\eta^2)$ at $\Gamma_s = \eta^2 \Gamma_m$ is inversely proportional to optical loss, scaling more favorably at high transmission so that $C \approx 10$ can be reached for $\eta^2 = 0.9$.

Our results demonstrate a comprehensive and versatile toolbox for generating coherent long-distance interactions with light and open up a range of exciting opportunities for quantum information processing, simulation and metrology. 
The coupling scheme constitutes a coherent feedback network \cite{Zhang2017} that allows quantum systems to directly exchange, process and feed back information without the use of classical channels.
The ability to create coherent Hamiltonian links between separate and physically distinct systems in a reconfigurable way significantly extends the available toolbox, not only for hybrid spin-mechanical interfaces \cite{Kurizki2015,Moller2017} but quantum networks \cite{Kimble2008} in general. It facilitates the faithful processing of quantum information and the generation of entanglement between spatially separated quantum processors across a room temperature environment.

\bibliographystyle{apsrev}
\bibliography{new_bib}


\section*{Acknowledgments}

We are grateful to Gianni Buser for setting up the dipole trap and acknowledge discussions with Maryse Ernzer.

\section*{Funding}

This work was supported by the project ``Modular mechanical-atomic quantum systems'' (MODULAR) of the European Research Council (ERC) and by the Swiss Nanoscience Institute (SNI). KH acknowledges support through the cluster of excellence ``Quantum Frontiers'' and from DFG through CRC 1227 DQ-mat,
projects A06.

\section*{Authors contributions}
TMK, BG, KH and PT conceived the experiment. TMK and BG developed the theory, with input from KH and PT. TMK, BG, CTN and GLS built the experimental setup, and TMK took and analyzed the data, discussing with PT. TMK, PT and KH wrote the manuscript with input from other authors. KH and PT supervised the project.

\section*{Competing interests}
The authors declare no competing interests.


\beginsupplement

\section*{Supplementary Materials for\\
Light-mediated strong coupling between a mechanical oscillator \\and atomic spins one meter apart}

\tableofcontents

\let\addcontentsline\oldaddcontentsline

\section{Details of the implementation}
\label{sec_suppl_1}

\subsection{Atomic spin ensemble}

\begin{figure}[b!]
\centering
\includegraphics[width = \linewidth]{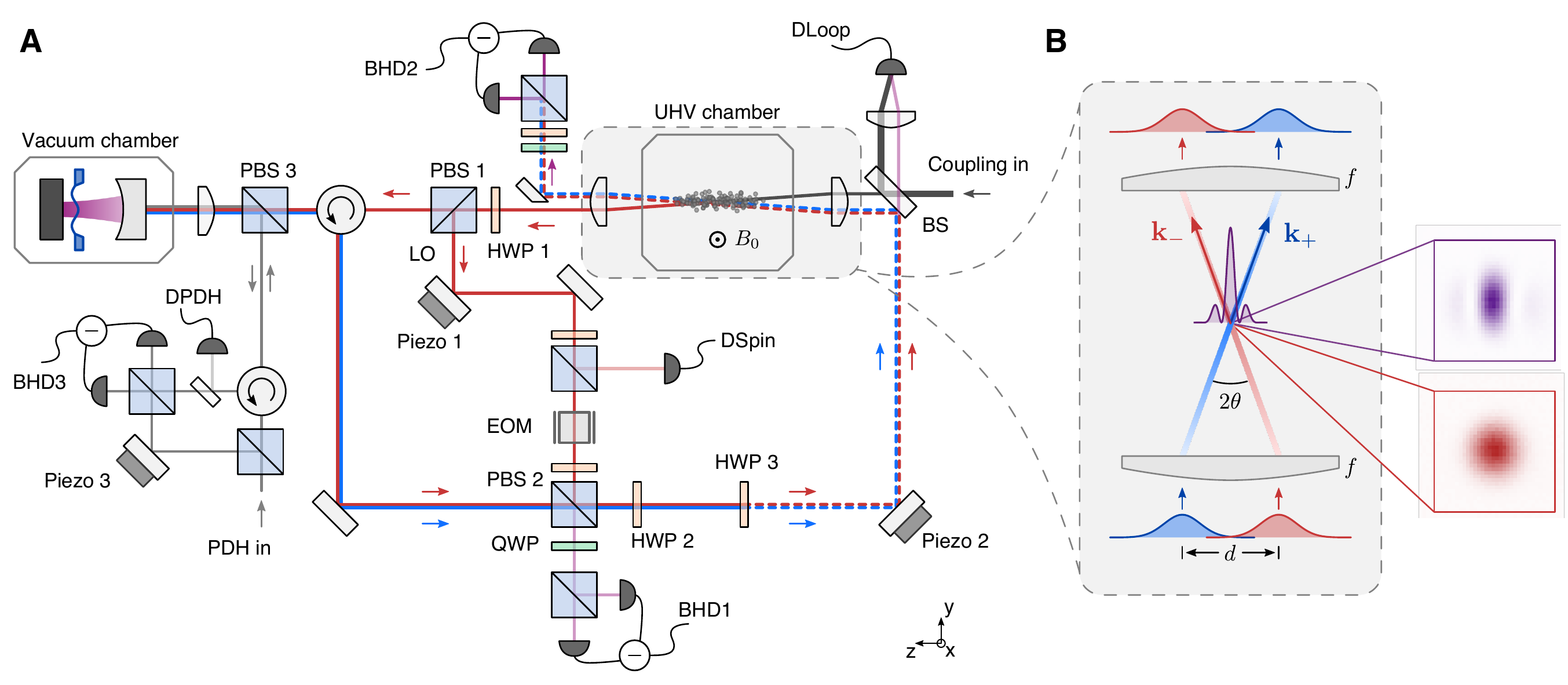}
\caption{Details of the experimental setup. (\textbf{A}), Sketch of the experiment showing optical path including relevant polarization optics, detectors and piezo-actuated mirrors for laser phase control. (\textbf{B}), Magnified view of the double-pass atom-light interface. Insets show measured beam profiles of coupling beam in the first pass (red) and of the interfering beams (purple).}
\label{fig_si_exp_setup_details}
\end{figure}

\paragraph{Experimental setup}
A detailed drawing of the full experimental setup is shown in Fig.~\ref{fig_si_exp_setup_details}. An ultracold cloud of $N \approx 10^7$ 87-Rubidium atoms is prepared at a temperature of $50~\mu$K in an optical dipole trap \cite{Grimm2000}, loaded from a magneto-optical trap within 1.2~$s$. The dipole trap is formed by a far off-resonant laser beam at $1064$~nm with optical power of $16$~W focused to a waist of $90~\mu$m. The resulting pencil-shaped atomic cloud has a $1/e^2$ diameter of $2 w_a = 60~\mu$m and length of $7$~mm. After the loading is completed, a constant magnetic field of $B_0 = \pm 2.8$~G is then applied transverse to the trap axis along the $x$-direction. The spin state is prepared by optically pumping all atoms to the $|f=2, m_f = \mp2\rangle$ hyperfine sublevel of the ${}^{2}S_{1/2}$ ground state using $\sigma_{\mp}$-polarized light with respect to the magnetic field. Optical pumping has an overall efficiency of about 90\% for $B_0=+2.8$~G and about 70\% for $B_0=-2.8$~G. In the frame relative to the magnetic field, spin precession is described by the Hamiltonian
\begin{equation}
H_0 = \hbar \gamma_f |B_0| F_x
\end{equation}
where $\gamma_f$ is the gyromagnetic ratio. For the strongly polarized spin ensemble with $|\bar{F}_x| \approx f N$, we can make a Holstein-Primakoff approximation such that $F_x \approx \bar{F}_x - \sgn(\bar{F}_x) (X_s^2 + P_s^2)/2$. This realizes a positive-mass oscillator for $\bar{F}_x<0$ or a negative-mass oscillator for $\bar{F}_x>0$ with oscillation frequency $\Omega_s = -\sgn(\bar{F}_x) \gamma_f |B_0|$ \cite{Polzik2015}.

The coupling laser is red-detuned by $\Delta_a = - 2\pi\times 80$~GHz from the Rubidium $D_2$ transition (${}^{2}S_{1/2}\to{}^{2}P_{3/2}$) at a wavelength of $\lambda = 780.241$~nm (laser frequency $\omega_L = 2\pi\times 384.148$~THz). Its linear polarization is adjusted to have an angle of 55$^\circ$ relative to the magnetic field in order to minimise spin dephasing due to inhomogeneous tensor light shifts \cite{Smith2004}. For good mode-matching with the atomic cloud, the laser is focused to a waist of $w_0 = 35~\mu$m with a corresponding Rayleigh length of $z_0=4.9~$mm. In the absence of coupling light, the intrinsic spin decoherence rate $\gamma_{s,0} = 2\pi \times 100$~Hz is limited by residual magnetic field inhomogeneities. Switching on the coupling light in single-pass increases the the spin damping rate to $\gamma_s \approx 2\pi\times 0.5$~kHz for an optical input power of $\hbar\omega_L \Phi_L=1$~mW at the detuning of $\Delta_a = -2\pi \times 80$~GHz. In the double-pass configuration used in the experiment, this value increases to $\gamma_s = 2\pi\times 4$~kHz, which is more than the expected four-fold increase due to the larger optical intensity. This excess damping rate could be explained by additional inhomogeneous broadening due to feedback via the loop, caused by imperfect polarization adjustment.

\paragraph{Single-pass interaction}
The interaction between the collective atomic spin $\mathbf{F} = \sum_{i=1}^{N} \mathbf{f}^{(i)}$ and the detuned laser field is well described by the Faraday interaction \cite{Hammerer2010}. For a single laser beam propagating along the $z$ direction, the corresponding interaction Hamiltonian reads
\begin{equation}
H_s = \hbar \alpha_1 F_z S_z
\end{equation}
where $\alpha_1$ is the dimensionless atomic vector polarisability and $S_z$ is the $z$ component of the Stokes vector describing the polarization state of the light field. The Stokes vector components
\begin{eqnarray}
S_0 &=& \frac{1}{2} (a_x^\dag a_x + a_y^\dag a_y),\nonumber\\
S_x &=& \frac{1}{2} (a_x^\dag a_x - a_y^\dag a_y),\nonumber\\
S_y &=& \frac{1}{2} (a_x^\dag a_y + a_y^\dag a_x),\nonumber\\
S_z &=& \frac{1}{2i} (a_x^\dag a_y - a_y^\dag a_x) \label{eq_si_stokes_vector_def}
\end{eqnarray}
have units of s$^{-1}$ and describe the total photon flux, the difference in photon flux between $x$- and $y$-polarization, the in-phase coherence between $x$- and $y$-polarization (polarization at $\pm45^\circ$) and the out-of-phase coherence of $x$- and $y$-polarization (circular polarization), respectively. Their commutation relations are $[S_k(z), S_l(z')] = i\epsilon_{klm} c \delta(z-z') S_l(z)$ ($k,l,m \in \{x,y,z\}$) with $c$ being the speed of light. For a strong, $x$-polarized laser field with photon flux $\Phi_L$, one can make a Holstein-Primakoff approximation for the Stokes vector resulting in $S_y = \sqrt{\Phi_L/2} X_L$ and $S_z = \sqrt{\Phi_L/2} P_L$. Here, $X_L = (a_y + a_y^\dag)/\sqrt{2}$ and $P_L = -i(a_y - a_y^\dag)/\sqrt{2}$ are the amplitude and phase quadratures, respectively, of the light mode in $y$-polarization, i.e. orthogonal to the laser field. This picture does not change when the laser polarization is rotated about the $z$-axis, as is the case in the experiment, one simply needs to redefine the $S_x$ and $S_y$ axes.

The vector polarisability for the $D_2$ line of Rubidium 87 at large detuning is given by $\alpha_1 = \frac{\lambda^2}{8\pi A}\frac{\gamma_\mathrm{se}}{\Delta_a}$, where $A = \pi w_0^2$ is the cross section of the laser beam, $\gamma_\mathrm{se} = 2\pi\times6.1$~MHz is the spontaneous emission rate of the excited ${}^{2}P_{3/2}$ state. For the present setup, the effective vector polarisability has been measured via Faraday rotation to amount to $\alpha_1 \approx 1.36(1)\times 10^{-8} (2\pi \mathrm{GHz}/\Delta_a)$. At the large detuning of $\Delta_a = -2\pi\times 80$~GHz, atom-light coupling via the atomic tensor polarisability is negligible.
 
In the experiment, optical pumping strongly polarises the collective spin along a transverse magnetic field with strength $B_0$ along the $x$-direction. We then have $\bar{F}_x \approx \sgn(\bar{F}_x) f N$, where $f=2$ is the spin per atom. In this case the spin dynamics are well approximated by a harmonic oscillator with quadratures $X_s = F_z / \sqrt{|\bar{F}_x|}$ and $P_s = - \sgn(\bar{F}_x) F_y /\sqrt{|\bar{F}_x|}$. They satisfy the canonical commutation relation $[X_s, P_s] = i$. The atom-light interaction Hamiltonian can now be written in the form
\begin{equation}
H_s = \hbar 2\sqrt{\Gamma_s} X_s P_L
\end{equation}
with spin measurement rate being defined as $\Gamma_s = \alpha_1^2 |\bar{F}_x| |\bar{S}_x|/4 = \alpha_1^2 N f \Phi_L/8$. The corresponding input-output relation for the light field interacting with the atoms is
\begin{eqnarray}
X_L^\mathrm{(out)} &=& X_L^\mathrm{(in)} + 2\sqrt{\Gamma_s} X_s\\
P_L^\mathrm{(out)} &=& P_L^\mathrm{(in)}
\end{eqnarray}

The average decoherence rate due to far-detuned spontaneous photon scattering induced by a linearly polarized laser beam is given by $\gamma_\mathrm{sc} = \frac{\lambda^2}{4\pi A}(\frac{\gamma_\mathrm{se}}{\Delta_{a}})^2 \Phi_L$. The single-pass atomic cooperativity (for $f=2$) is given by the ratio
\begin{equation}
c_s = \frac{4\Gamma_s}{\gamma_\mathrm{sc}} = \frac{N \lambda^2}{16\pi A} = \frac{d_0}{16}
\end{equation}
which has been expressed in terms of the resonant optical depth $d_0$ for linearly polarized light with cross-section $\lambda^2/\pi$. The total spin damping rate $\gamma_s = \gamma_{s,0} + \gamma_\mathrm{sc}$ is the sum of the intrinsic damping rate and spontaneous scattering rate.

\paragraph{Geometrical considerations for double-pass interaction}

In order to establish the bidirectional spin-membrane coupling, the laser beam returning from the optomechanical system is sent another time through the atomic cloud. A small angle $2\theta\ll 1$ between the first and second pass allows us to fully separate them after passing through the atomic ensemble while still maintaining good alignment with the atomic cloud and high optical depth (see Fig.~\ref{fig_si_exp_setup_details}B). The two beams have wave-vectors $\mathbf{k}_\pm$ with angle $\pm\theta$ relative to the $z$-axis, respectively.

The two beams are initially aligned parallel with displacement $d$ and then focused onto the atomic cloud using a lens with focal length $f=200$~mm which transforms the parallel offset into a relative angle $\theta \approx d/2f$. In the focal plane, the electric fields of the two beams interfere and form a transverse standing wave with effective wavelength $\lambda_\perp \approx \lambda/2\theta$. The global phase difference between the two beams is stabilized to zero (constructive interference at the center) as explained in section~\ref{sec_suppl_2}. Camera images of the single-pass and double-pass laser beam profiles at the position of the atoms are shown in Fig.~\ref{fig_si_exp_setup_details}B. These measured by picking up the $\approx 0.1\%$ transmission through one of the mirrors which align the beams onto the atomic cloud.

In order to ensure that both beams couple to the same atomic spin wave, it is crucial that the transverse wavelength is equal to or larger than the diameter of the atomic cloud $2w_a$. In order to optimise the mode-matching between the laser and the atomic cloud in single pass, the laser waist has been chosen to approximately match that of the atomic cloud, i.e. $w_0 \approx w_a$. In this situation we require that $\lambda_\perp / 2 w_0 \approx \pi \theta_0/4\theta > 1$, where $\theta_0 = \lambda/\pi w_0$ is the beam divergence. On the other hand, the ability to separate the two beams in the plane of the lens requires that the beam separation is larger than the collimated beam waist $w$, i.e. $d / 2 w = \theta/\theta_0 > 1$. For these two conflicting requirements we find a compromise by choosing $\theta \approx \theta_0$, i.e. a displacement $d \approx 2 w$. For this setting, the residual beam overlap in the lens plane is approximately $10\%$. Choosing a larger beam diameter in the focus would allow even smaller angles with better homogeneity on the atomic cloud. There is however another tradeoff with the mode-matching efficiency of the laser beam to the atomic ensemble. 

The presence of two laser fields, changes both the spontaneous scattering rate $\gamma_\mathrm{sc}$ and the single-pass spin-light coupling strength $\Gamma_s$. Since there are now two laser fields of equal flux and with same polarization, the total intensity increases by four and so does $\gamma_\mathrm{sc}$. Since the two beams are nearly collinear, their combined fields can achieve a higher coherent scattering rates into the forward modes going towards the optomechanical system and towards the output. Likewise, optical signals from the optomechanical system couple more strongly to the spin because the pump strength is larger. This results in an enhancement of $\Gamma_s$ which is, however, smaller than that of $\gamma_\mathrm{sc}$. The reason for this is that spontaneous scattering is a single-atom effect, while collective forward scattering relies on the constructive interference of all fields scattered by the individual atoms \cite{Baragiola2014}. The small angle between the two laser fields results in a transverse phase pattern $\sim 1 + e^{\pm i 2 k_\perp y}$ of the single-atom scattering amplitude into the modes with wave-vectors $\mathbf{k}_{\pm}$. Here, the transverse wave-number is defined as $k_\perp = 2\pi/\lambda_\perp$. Consequently, atoms at different transverse locations in the laser beam scatter light with different relative phases which reduces the collective enhancement of forward scattering. Furthermore, this phase-pattern results in different spin waves $F_z^{1 + \cos{}}$ and $F_z^{\sin{}}$ with local amplitudes $1 + \cos(2k_\perp y)$ and $\sin(2k_\perp y)$, respectively \cite{Vogell2015}. In order to achieve that only the homogeneous spin-wave $F_z^{1+\cos{}}$ has a large coupling strength, one must ensure that the atoms are sufficiently localized such that $k_\perp w_a  \ll 1$. Since we have $\theta \approx \theta_0$, this condition requires $w_a\ll w_0$, ie. tight transverse confinement of the atomic ensemble. In the experiment, this condition is not satisfied such that also the spin wave $F_z^{\sin{}}$ has a non-negligible coupling strength. This has no consequence for coherent coupling as explored in this article, but coupling between the two spin-waves can act as a source of noise.

\subsection{Optomechanical system}

Our optomechanical system consists of a $100$~nm thin square silicon-nitride membrane (side length $400~\mu$m) mounted inside an optical cavity. The silicon-nitride membrane is supported by a silicon chip with phononic band gap structure to shield the target mode from acoustic noise of the silicon frame \cite{Yu2014}. We work with the 2,2-mode at $\Omega_m = 2\pi\times1.9576$~MHz which has a quality factor of $1.3\times10^6$ at room temperature and an effective mass of $m_\mathrm{eff} = 12$~ng. The membrane is mounted about $100~\mu$m from the flat, high-reflectivity end mirror of a 1.2~mm long, single-sided optical cavity.The optomechanical cavity assembly is mounted in vacuum in a room temperature environment. At the working point, the cavity linewidth is $\kappa = 2\pi\times63$~MHz and the vacuum optomechanical coupling constant is $g_0=2\pi\times 220$~Hz for the 2,2-mode. This places our optomechanical system deep in the non-resolved sideband regime, such that coupling light in and out of the cavity occurs on a time scale that is much faster than the mechanical oscillation period. The coupling light enters through the partially reflecting end mirror of the cavity with a coupling efficiency of $\eta_c \approx0.9$, leading to a cavity power reflectivity on resonance of $(1-2\eta_c)^2 \approx 0.6$. Spatial mode-matching between the input laser beam and the cavity mode is achieved with an efficiency $>90\%$. The polarization interferometer mode-matching efficiency at the output PBS is also optimized to $>90\%$. 

The optical cavity is continuously locked by an independent laser beam which is 5~MHz red detuned from the coupling laser. The lock point is adjusted such that the coupling beam is only slightly red detuned from the cavity resonance to avoid optomechanical instabilities. Dynamical back-action from the combined intra-cavity field increases the mechanical damping rate to $\gamma_m = 2\pi\times 300$~Hz and cools its motion to $\bar{n}_m \approx 1.5\times 10^4$ phonons. Homodyne detection of the cavity lock-beam reflected from the cavity serves as an independent measurement of the membrane motion. Part of the electronic detector signal is used for feedback cooling of the 1,1 mechanical mode, whose second harmonic would otherwise appear near the 2,2-mode and disturb the experiment.

Our membrane-at-the-end cavity can be described by the canonical cavity optomechanics framework \cite{Aspelmeyer2014}. Following standard optomechanical theory we write the Hamiltonian for the optomechanical system in a rotating frame at the laser frequency $\omega_L$
\begin{eqnarray}
H &=& H_0 + H_\mathrm{om}\\
H_0 &=& - \hbar \Delta_c c^\dag c + \frac{\hbar\Omega_m}{2} (X_m^2 + P_m^2)\\
H_\mathrm{om} &=& \hbar  \sqrt{2} g_0 \;c^\dag c \; X_m \label{eq_optomech_interaction}
\end{eqnarray}
Here, $c$ is the annihilation operator of the cavity field and $\Delta_c$ is the laser-cavity detuning. The mechanical mode has annihilation and creation operators $b_m$ and $b_m^\dag$, respectively. The mechanical position and momentum quadratures (normalized in units of the zero-point motion $x_\mathrm{zpf} = \sqrt{\hbar/2m_\mathrm{eff}\Omega_m}$) are $X_m = (b_m + b_m^\dag)/\sqrt{2}$ and $P_m = (b_m - b_m^\dag)/\sqrt{2}i$, respectively, satisfying the canonical commutation relation $[X_m,P_m] = i$. Coupling of the cavity  mode to the external traveling field (ignoring coupling to other input modes) is described by
\begin{equation}
H_\mathrm{ext} = i \hbar\sqrt{\eta_c\kappa} (a_L^\dag(\zeta_m) c - c^\dag a_L(\zeta_m))
\label{eq_H_ext_cavity}
\end{equation}
where $\zeta_m$ denotes the coordinate of the optomechanical system along the optical path of $a_L$. 

The optomechanical interaction linearises in the presence of a large coherent state amplitude of the cavity field $\alpha_c = \sqrt{\eta_c\kappa\Phi_L}|\chi_c(0)|$, resulting from an external drive with photon flux $\Phi_L$ via the external field $a_L$. Here, the cavity susceptibility $\chi_c(\omega) = (\kappa/2 - i (\Delta_c + \omega))^{-1}$ has been defined. The linearized Hamiltonian reads
\begin{equation}
H_\mathrm{om} = \hbar  \sqrt{2}g_\mathrm{om}\left(e^{-i\phi_c} c + e^{i\phi_c} c^\dag\right) X_m,
\label{eq:H_om_linear}
\end{equation}
with coherently enhanced optomechanical coupling strength $g_\mathrm{om}=g_0\alpha_c$. The phase shift between cavity field and external field is $\phi_c = \arg\chi_c(0) = \arctan(2\Delta_c/\kappa)$. In the non-resolved sideband limit, $\kappa \gg \Omega_m$, the cavity field can be adiabatically eliminated resulting in an effective interaction between the mechanical oscillator and the external optical mode. We  start by writing the equation of motion for the cavity
\begin{equation}
\dot{c} = \left(i \Delta_c - \frac{\kappa}{2}\right) c - \sqrt{\eta_c\kappa} a_\mathrm{in} - i \sqrt{2}g_\mathrm{om} e^{i\phi_c} X_m
\end{equation}

In the non-resolved sideband limit and on cavity resonance ($\Delta_c = 0$) one can neglect the phase shifts associated with detuning or sideband resolution and finds a steady-state cavity amplitude $c \approx -\sqrt{\eta_c\kappa} a_\mathrm{in}/(\kappa/2 - i \Delta_c)$ (omitting the mechanical modulation of the cavity field for simplicity). This expression can be inserted into \eqref{eq:H_om_linear} such that we get the effective optomechanical coupling to the external field 
\begin{equation}
H_{m} = \hbar 2\sqrt{\Gamma_m} X_m X_L(\zeta_m)
\label{eq:om_Hamiltonian_eliminated_simple}
\end{equation}
with optomechanical measurement rate defined as 
\begin{equation}
\Gamma_m = \eta_c \frac{4 g_\mathrm{om}^2}{\kappa} = \eta_c^2 \left(\frac{4 g_0}{\kappa}\right)^2 \Phi_L
\end{equation}

The resulting input-output relations for the light field at position $\zeta > \zeta_m$ read
\begin{eqnarray}
X_L^\mathrm{(out)}(\zeta,t) &=& X_L^\mathrm{(in)}(\zeta,t)\\
P_L^\mathrm{(out)}(\zeta,t) &=& P_L^\mathrm{(in)}(\zeta,t) - 2\sqrt{\Gamma_m} X_m(t- (\zeta-\zeta_m)/c).
\end{eqnarray}

\subsection{Optical interface}

\begin{figure}[b!]
\centering
\includegraphics[width = 0.66\linewidth]{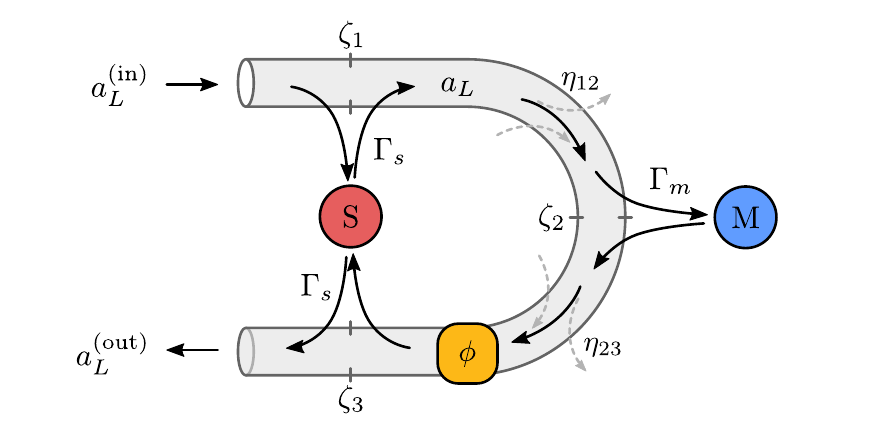}
\caption{Sketch of the cascaded coupling. The optical field $a_L$ takes a path parametrized by a spatial coordinate $\zeta$ from the spin system S to the membrane M and back to S. The field experiences loss between the systems, characterized by transmission coefficients $\eta_{ij}$ and a phase shift $\phi$, i.e. the loop phase.}
\label{fig_si_cascade_1}
\end{figure}

In this section, we describe in detail how the optical setup is designed to mediate an interaction between the spin and optomechanical systems. We derive an abstract description of the setup in the language of cascaded quantum systems, which is sketched in Fig.~\ref{fig_si_cascade_1}.

Outside the polarization interferometer, we describe the polarization state of light using the Stokes vector $\mathbf{S}$ (see equations~\eqref{eq_si_stokes_vector_def}). At the input, the laser is linearly polarized along the $x$-axis such that $\bar{S}_x = \bar{S}_0 = \Phi_L/2$. The field amplitudes read $a_x = \sqrt{\Phi_L} + b_L$ in $x$-polarization and $a_y = a_L$ in $y$-polarization, where we have defined $a_L$ and $b_L$ as the quantum fields in $y$- and $x$-polarization, respectively. Then, assuming $\langle a_L^\dag a_L\rangle, \langle b_L^\dag b_L\rangle \ll \Phi_L$, the Stokes vector components can be linearized and written as
\begin{eqnarray}
S_0 &\approx& \bar{S}_0 + \sqrt{\frac{\bar{S}_0}{2}}(b_L + b_L^\dag),\nonumber\\
S_x &\approx& \bar{S}_0 + \sqrt{\frac{\bar{S}_0}{2}}(b_L + b_L^\dag),\nonumber\\
S_y &\approx& \phantom{-i}\sqrt{\frac{\bar{S}_0}{2}}(a_L + a_L^\dag),\nonumber\\
S_z &\approx& - i \sqrt{\frac{\bar{S}_0}{2}}(a_L - a_L^\dag).\label{eq_stokes_linearization}
\end{eqnarray}

The spin-light interaction in the first pass, at optical path coordinate $\zeta_1$, reads
\begin{equation}
H_{s,1} = 2 \hbar \sqrt{\Gamma_s/\bar{S}_x} X_s S_z(\zeta_1) \approx 2 \hbar \sqrt{\Gamma_s} X_s P_L(\zeta_1)
\end{equation}
where $P_L = -i(a_L - a_L^\dag)/\sqrt{2}$ is the phase quadrature of the $y$-polarized quantum field. Likewise, $X_L = (a_L + a_L^\dag)/\sqrt{2}$ is the amplitude quadrature. Spin precession modulates the light polarization via the input-output relation
\begin{equation}
S_y^\mathrm{(out),1} = S_y^\mathrm{(in),1} + 2\sqrt{\Gamma_s \bar{S}_0} X_s
\end{equation}
Before entering the polarization interferometer, the laser polarization is rotated by a half-wave plate at angle $\theta_H$. This transforms the Stokes vector as 
\begin{eqnarray}
S_{x}' &=& \phantom{+} S_{x} \cos(4\theta_H) + S_{y} \sin(4\theta_H)\\
S_{y}' &=& - S_{y} \cos(4\theta_H) + S_{x} \sin(4\theta_H)\\
S_{z}' &=& - S_{z}
\end{eqnarray}
The photon flux in the interferometer arm containing the optomechanical cavity is given by
\begin{equation}
a_{y}'^\dag a'_{y} = S_0 - S_{x}' = S_0 - \cos(4\theta_H) S_x - S_y \sin(4\theta_H)
\end{equation}
In the limit of a broad cavity linewidth, the optomechanical interaction \eqref{eq_optomech_interaction} can be written as
\begin{equation}
H_{m} = \hbar \frac{4 g_0}{\kappa} \sqrt{2}X_m a_{y}'^\dag a_{y}'
\end{equation}
where we have made the substitution $c^\dag c = (4/\kappa) a_{y}'^\dag a_{y}'$ of the cavity photon number by the input photon flux. Linearizing about the strong laser field (using equations \eqref{eq_stokes_linearization}) yields
\begin{equation}
H_{m} = \hbar \frac{4 g_0}{\kappa} \sqrt{\bar{S}_0} X_m \left[(b_L + b_L^\dag)(1-\cos(4\theta_H)) - \sin(4\theta_H) (a_L + a_L^\dag)\right]
\end{equation}
The first term describes coupling to the $x$-polarized quantum field co-propagating with the laser. In this context, it must be interpreted as noise because the spin does not interact with it. The second term is the $y$-polarized quantum field which contains the spin signal and is relevant for the cascaded coupling. In order to couple the mechanical oscillator mostly to $a_L$ and not $b_L$, we choose a small half-wave plate angle $\theta_H \ll 1$ such that only about $ 1-\cos(4\theta_H)= 0.1$ of the laser light is transmitted towards the optomechanical cavity. This still results in a large value of $\sin(4\theta_H) \approx 0.5$ while the ratio of back-action due to $a_L$ over the total back-action of $a_L$ and $b_L$, $\sin(4\theta_H)^2/([1-\cos(4\theta_H)]^2 + \sin(4\theta_H)^2) \approx 0.93$, is high.

The optomechanical measurement rate is then given by $\Gamma_m = (4g_0/\kappa)^2 \bar{S}_0 \sin(4\theta_H)^2/2 = (4g_0/\kappa)^2 \Phi_m$ with effective photon flux $\Phi_m = \Phi_L \sin(4\theta_H)^2/4$ at the optomechanical cavity. The spin-induced amplitude modulation at the optomechanical cavity amounts to $2 \sin(4\theta_H)\sqrt{\Gamma_s \bar{S}_0} X_s$.

Mechanical motion produces a phase-modulation of the cavity output field by an angle $\phi_m = (4 g_0/\kappa) \sqrt{2}X_m$. This phase shift between the two interferometer arms maps onto the output Stokes vector as
\begin{eqnarray}
{S_{x}'}^\mathrm{(out)} &=& {S_{x}'}^\mathrm{(in)} \\
{S_{y}'}^\mathrm{(out)} &=& {S_{y}'}^\mathrm{(in)} \cos(\phi_m) + {S_{z}'}^\mathrm{(in)}\sin(\phi_m)\\
{S_{z}'}^\mathrm{(out)} &=& {S_{z}'}^\mathrm{(out)} \cos(\phi_m) - {S_{y}'}^\mathrm{(in)}\sin(\phi_m)
\end{eqnarray}
Since $|\phi_m| \ll 1$ and $\langle {S_{z}'}^\mathrm{(in)}\rangle = 0$, $\langle {S_{y}'}^\mathrm{(in)}\rangle = \bar{S}_0 \sin(4\theta_H)$ we obtain that the polarization modulation due to the membrane at the second atom-light interaction amounts to $S_z = 2\sqrt{\bar{S}_0\Gamma_m} X_m$. 

The loop phase can be tuned by placing additional wave plates in the optical path before the second atom-light interaction. For the experiments presented in the main text, we used a single half-wave plate with fast axis aligned parallel to the laser polarization along $x$. This retards the orthogonal $y$-polarization by $\phi = \pi$ and thus inverts both $S_y$ and $S_z$. A continuous rotation of the Stokes vector about the $S_x$ axis by an angle $\phi\in[0,2\pi)$ can be performed using a stack of two quarter-wave plates (QWP) and one half-wave plate (HWP) in between. This requires aligning the fast axes of the QWP at $45^\circ$ ($\pi/4$) relative to the $x$-axis. For a rotation angle $\phi/4$ of the HWP relative to the QWP axes we obtain
\begin{equation}
\left[\mathrm{QWP}(\frac{\pi}{4})\circ \mathrm{HWP}(\frac{\pi + \phi}{4}) \circ \mathrm{QWP}(\frac{\pi}{4})\right] \mathbf{S} = \begin{pmatrix}
S_x\\ S_y \cos(\phi) - S_z \sin(\phi)\\ S_z \cos(\phi) +S_y \sin(\phi)
\end{pmatrix}
\end{equation}
which is the desired rotation about the $S_x$ axis. We remark that one can perform two such phase rotations, in between subsequent light-matter interfaces, to implement arbitrary couplings.

To summarize, we can describe the experimental setup of Fig.~\ref{fig_si_exp_setup_details} by cascaded light-matter interactions as depicted in a more abstract way in Fig.~\ref{fig_si_cascade_1}. We can write the cascaded interaction Hamiltonian with the traveling quantum field $a_L$ as
\begin{equation}
H_\mathrm{int} = 2\hbar \sqrt{\Gamma_s} X_s \left[P_L(\zeta_1) + \cos(\phi) P_L(\zeta_3) - \sin(\phi) X_L(\zeta_3)\right] + 2\hbar \sqrt{\Gamma_m} X_m X_L(\zeta_2)
\end{equation}
where $\zeta_1 < \zeta_2 < \zeta_3$ are the spatial coordinates of the three light-matter interactions along the optical path. This interaction Hamiltonian is the starting point to derive a master equation for the effective light-mediated coupling based on the formalism of ref.~\cite{Karg2019}. In the course of adiabatic elimination of the light field we drop propagation delays $\tau_{ij} = (\zeta_i - \zeta_j)/c$, which can be accounted for using Heisenberg-Langevin equations as presented in section \ref{sec_hle}. For the moment we also neglect optical loss. The resulting master equation is
\begin{eqnarray}
\dot{\rho} 	&=& - 2\Gamma_s (1 + e^{-i\phi}) [X_s, X_s \rho] + \mathrm{h.c.}\\
			& &- \Gamma_m  [X_m, X_m \rho] + \mathrm{h.c.}\\
			&& + 2 i \sqrt{\Gamma_s \Gamma_m} \left( [X_m, X_s \rho] - e^{-i\phi} [X_s, X_m \rho]\right) + \mathrm{h.c.} \label{eq_master_eq_spinmembrane_raw}
\end{eqnarray}
where the first line contains spin diffusion due to vacuum noise of the optical input field and light-mediated spin self-interaction. The second line corresponds to mechanical diffusion due to optical input noise. Spin-membrane interaction, both coherent and dissipative, is contained in the third line. We can separate coherent from dissipative evolution by bringing the master equation~\eqref{eq_master_eq_spinmembrane_raw} into Lindblad form, i.e.
\begin{equation}
\dot{\rho} = \frac{1}{i\hbar} [H_\mathrm{eff}, \rho] + \mathcal{L}_\mathrm{eff} \rho,
\end{equation}
with an effective Hamiltonian
\begin{equation}
H_\mathrm{eff} = \hbar [1-\cos(\phi)]2\sqrt{\Gamma_s \Gamma_m} X_s X_m - \hbar \sin(\phi)2\Gamma_s X_s^2
\end{equation}
and collective dissipation
\begin{equation}
\mathcal{L}_\mathrm{eff} \rho = \mathcal{D}[J]\rho = J \rho J^\dag -\frac{1}{2} \left(J^\dag J \rho + \rho J^\dag J\right)
\end{equation}
with collective jump operator $J = \sqrt{2 \Gamma_m} X_m + i \sqrt{2 \Gamma_s}\left(1 + e^{i\phi}\right)X_s$.

The effective Hamiltonian $H_\mathrm{eff}$ contains both the spin-membrane interaction and a self-interaction of the spin. For loop phases $\phi = 0, \pi$, the spin self-interaction vanishes and is thus not important for the experiment. At intermediate phases, the spin self-interaction can be exploited to generate unconditional spin-squeezing \cite{Takeuchi2005}. The spin-membrane interaction vanishes for $\phi = 0$ and amounts to $H_\mathrm{eff} = \hbar 2 g X_s X_m$ for $\phi = \pi$. Here, we define the spin-membrane coupling strength as $g = 2\sqrt{\Gamma_s \Gamma_m}$. The collective jump operator $J$ is composed of a membrane and a spin part. At $\phi = 0$, both parts are non-zero and give rise to collective dissipative interaction between spin and membrane. For $\phi = \pi$, however, the spin part vanishes, and dissipation only affects the membrane. This is a consequence of the destructive interference of optical shot noise driving the spin for $\phi = \pi$, in which case the spin is effectively decoupled from the input and output light fields. Quantum-coherent spin-membrane coupling necessitates this property of the looped cascaded coupling scheme in order to allow the coupling strength $g$ to be larger than all back-action decoherence rates. If intrinsic dissipation rates are low, this is achieved if $\Gamma_s > \Gamma_m$ because $g/\Gamma_m = \sqrt{\Gamma_s/\Gamma_m}$. Including thermal decoherence of the individual oscillators, the full master equation reads
\begin{eqnarray}
\dot{\rho} 	&=& \frac{1}{i\hbar}[H_0 + H_\mathrm{eff}, \rho] + \mathcal{L}_\mathrm{eff}\rho\\
			& & + \gamma_m (\bar{n}_m + 1) \mathcal{D}[b_m] \rho + \gamma_m \bar{n}_m \mathcal{D}[b_m^\dag] \rho\\
			& & + \gamma_s (\bar{n}_s + 1) \mathcal{D}[b_s] \rho + \gamma_s \bar{n}_s \mathcal{D}[b_s^\dag] \rho
\end{eqnarray}
Here, we introduced the harmonic oscillator Hamiltonian
\begin{equation}
H_0 = \frac{\hbar \Omega_m}{2} (X_m^2 + P_m^2) + \frac{\hbar \Omega_s}{2} (X_s^2 + P_s^2)
\end{equation}
and intrinsic damping rates $\gamma_i$ with thermal bath occupation numbers $\bar{n}_i$ ($i = s, m$). For the spin system $\bar{n}_s \approx 0$.

Optical losses lead to a slight modification of the ideal effective dynamics derived above. We introduce the transmission coefficients $\eta_{ij}$ between light-matter couplings $i$ and $j$ (see Fig.~\ref{fig_si_cascade_1}). Since the laser field experiences the same loss as the quantum field that mediates the coupling we also need to scale the local coupling strengths $\Gamma_i$ as they are proportional to the local laser photon flux. The modified master equation with losses reads
\begin{eqnarray}
\dot{\rho} 	&=& \frac{1}{i\hbar}[H_0, \rho] + \sum_{i=s,m} \left(\gamma_{i} (\bar{n}_i + 1) \mathcal{D}[b_i] \rho + \gamma_{i} \bar{n}_i \mathcal{D}[b_i^\dag] \rho \right)\nonumber\\
			& & - \Gamma_s (1 + \eta_{13}^2 + 2 \eta_{13}^2e^{-i\phi}) [X_s, X_s \rho] + \mathrm{h.c.}\nonumber\\
			& &- \eta_{12}^2 \Gamma_m  [X_m, X_m \rho] + \mathrm{h.c.}\nonumber\\
			&& - 2 i \sqrt{\Gamma_s \Gamma_m} \left( \eta_{12}^2 [X_m, X_s \rho] -  \eta_{12}\eta_{23}\eta_{13} e^{-i\phi} [X_s, X_m \rho]\right) + \mathrm{h.c.} \label{eq_master_eq_spinmembrane_loss}
\end{eqnarray}
For identical transmission coefficients $\eta_{12} = \eta_{23} = \eta$ and $\eta_{13} = \eta^2$ and $\phi = \pi$, we get a coherent coupling strength of $g = (\eta^2 + \eta^4)\sqrt{\Gamma_s \Gamma_m}$, a mechanical back-action rate of $\gamma_{m,\mathrm{ba}} = \Gamma_m \eta^2$, and a spin back-action rate of $\gamma_{s,\mathrm{ba}} = \Gamma_s (1 - \eta^4)$. 

After the second atom-light interaction, the output field is given by
\begin{eqnarray}
a_L^{\mathrm{(out)}}(t) &=& \eta_{12} \eta_{23} a_L^{\mathrm{(in)}}(t) + \eta_{23} \sqrt{1-\eta_{12}^2} h_{1}^{\mathrm{(in)}}(t) + \sqrt{1-\eta_{23}^2}h_2^{\mathrm{(in)}}(t)\nonumber\\
	& & + 2\eta_{12} \eta_{23}\left[\sqrt{\Gamma_s} X_s(t) + \sqrt{\Gamma_s} X_s(t-2\tau)e^{i\phi} + i\sqrt{\Gamma_m} X_m(t-\tau)e^{i\phi} \right]
\label{eq_si_output_field}
\end{eqnarray}
Here, we defined vacuum fields $h_{1}^\mathrm{(in)}$ and $h_{2}^\mathrm{(in)}$ that enter the coupling light field due to losses on the spin-to-membrane path and membrane-to-spin path, respectively. Since losses affect not only the quantum field $a_L$, but also the laser field, full interference of the spin output signal can be observed, even if there is significant optical loss in between the two spin-light interactions.


\section{Experiment control}
\label{sec_suppl_2}

In this section we describe specifics of the experiment control system and signal processing. Timing of the experimental sequence is controlled using the \emph{labscript suite} \cite{Starkey2013}.

\paragraph{Polarization interferometer} At the output of the polarization interferometer, i.e. after PBS 2 in Fig.~\ref{fig_si_exp_setup_details}B, the two arms must interfere constructively. To lock the phase, we split about 3\% of the total power from each arm at PBS 2 and send it to the balanced homodyne detector BHD 1 which measures the relative phase fluctuations between the two beams. The DC part of the detector output is directly used as an error signal to lock the phase difference at BHD 1 to $\pi/2$ by controlling the position of piezo 1. To ensure that the main beam going towards the atomic setup has a phase difference of zero, we place a quarter-wave plate in front of BHD 1 that compensates the $\pi/2$ phase shift of the lock. We use a digital FPGA-based proportional-integral controller \cite{Neuhaus2017} which can be set on hold while the coupling beam is off. The coupling beam must for example be switched off during the dipole-trap loading and optical pumping sequence of the atoms.

\paragraph{Cavity lock} The optomechanical cavity is locked via the Pound-Drever-Hall technique using a separate laser beam. This lock beam is shifted in frequency by $-5$~MHz relative to the coupling beam to avoid interference at the mechanical frequency, and in addition provides some optomechanical cooling of the membrane modes by dynamical back-action. Hence, the lock point is adjusted such that the coupling beam is on cavity resonance, and the lock beam is red-detuned. The cavity lock beam is also used to detect the membrane motion by balanced homodyne detection of the reflected beam on BHD3.

\paragraph{Phase lock of the double-pass} The phase difference between the two laser beams passing through the atomic ensemble under an angle is stabilized such that they show constructive interference for maximal atom-light coupling (see Fig.~\ref{fig_si_exp_setup_details}B). To achieve this, we stabilize the phase between the light returning from the optomechanical system that is transmitted through the input beam splitter (BS, transmission 2\%) and the directly reflected input laser beam. These beams have a transverse displacement of about 2~mm and are first fiber-coupled into a single-mode polarization-maintaining fiber and then detected on the photodetector DLoop. In order to distill the phase information from the large DC signal, we weakly phase-modulate the coupling beam using an electro-optic modulator (EOM, transmission 98 \%) placed inside the reference arm of the polarization interferometer. Demodulating the AC part of the beat signal at the modulation frequency (about 700 kHz) generates an error signal which exhibits a zero-crossing for constructive interference of the two beams at the location of the atoms. The feedback loop is closed using another FPGA-based controller that controls the position of piezo 2. Identical to the lock of the polarization interferometer, this lock is paused whenever the coupling beam is switched off. 

\paragraph{Signal processing}

Signals from the balanced homodyne detector BHD3 measuring the membrane signal and from the direct detector DSpin measuring the spin signal are demodulated at a frequency $\Omega_0$ close to the mechanical frequency $\Omega_m$ using a digital lock-in amplifier (Zurich instruments HF2LI). The membrane and spin detector signals normalized to their respective local oscillator powers can be written as
\begin{equation}
D_i(t) = \beta_i X_i(t) + W_i(t)
\end{equation}
where $\beta_m = 8 g_0 / \kappa$ for the membrane and $\beta_s = \alpha_1 \sqrt{\bar{F}_x}$ for the spin. Detector noise is described by the term $W_i$ and includes both optical shot noise and electronic noise. The demodulator outputs the in-phase and quadrature components
\begin{eqnarray}
I_i &=& \sqrt{2}\langle D_i(t) \cos(\Omega_0 t)\rangle_t,\\
Q_i &=& \sqrt{2}\langle D_i(t) \sin(\Omega_0 t)\rangle_t,
\end{eqnarray}
respectively, where $\langle \cdot \rangle_t$ denotes temporal averaging with a bandwidth of $40$~kHz. We then define the following slowly varying position and momentum quadratures
\begin{eqnarray}
\tilde{X}_m'(t) &=& \frac{\sqrt{2}}{\beta_m} I_m(t),\\
\tilde{P}_m'(t) &=& \frac{\sqrt{2}}{\beta_m} Q_m(t),\\
\tilde{X}_s'(t) &=& \phantom{-}\frac{\sqrt{2}}{\beta_s} (\cos(\alpha) I_s(t) + \sin(\alpha) Q_s(t)),\\
\tilde{P}_s'(t) &=& \pm\frac{\sqrt{2}}{\beta_s} (\cos(\alpha) Q_s(t) - \sin(\alpha) I_s(t)).
\end{eqnarray}
In the last line $+$($-$) refers to the positive (negative) spin oscillation frequency. The local phase $\alpha$ for the spin quadratures is adjusted to a value of $\alpha = 100^\circ$ in post-processing to optimise the measured spin-membrane correlations for the parametric-gain interaction. 

To calculate the number of excitations in each oscillator we use the formula $\tilde{n}_i = (\tilde{X}_i^2 + \tilde{P}_i^2)/2$. Estimates of the symmetrized mechanical power spectral densities are calculated using a fast-Fourier-transform (FFT), i.e.
\begin{equation}
\bar{S}_{XX,m}(\omega) = \frac{1}{M_\mathrm{sa} f_\mathrm{sa}}|\mathrm{FFT}[I_m(t) + i Q_m(t)](\omega)|^2
\end{equation}
where $f_\mathrm{sa}$ is the sampling rate and $M_\mathrm{sa}$ is the number of samples of the measurement record.


\section{Interference in the double-pass spin-light interaction}
\label{sec_suppl_3}

In this section, we discuss the data showing interference of the spin signal in the coupling beam output field (Fig.~\ref{fig_5}E,F of the main text). This intends to show that the coupling scheme presented here is capable of suppressing optical back-action by the light field on the spin, since spin information is prevented from leaking to the environment \cite{Braginsky1980}. For this measurement, the optomechanical cavity is tuned off-resonant from the laser such that there is no coupling of the spin to the membrane. After optical pumping, a short (30 $\mu$s) RF-pulse at the Larmor frequency excites spin precession with a small amplitude. Immediately afterwards, the coupling beam is switched on and the spin-induced Faraday rotation is detected on a balanced homodyne detector (BHD 2 in Fig.~\ref{fig_si_exp_setup_details}A). The detector is adjusted such that it measures the $X_L$ quadrature of the output field given by equation~\eqref{eq_si_output_field}, i.e.
\begin{eqnarray}
X_L^{\mathrm{(out)}}(t) &\approx& \eta^2 X_L^{\mathrm{(in)}}(t) + \sqrt{1-\eta^4} X_h^{\mathrm{(in)}}(t) \nonumber\\
	& & + 2\eta^2 \sqrt{\Gamma_s} \left\{ X_s(t) \Big[ 1 +  \cos(\phi)\cos(2\Omega_s\tau)\Big] - P_s(t) \cos(\phi) \sin(2\Omega_s\tau)\right\}\quad
\label{eq_x_output_field}
\end{eqnarray}
The first line is shot noise from the input field $X_L^{\mathrm{(in)}}$ and an additional contribution $X_h^\mathrm{(in)}$ due to optical loss along the optical path. The second line contains the interfering spin signal. Here, we approximated $X_s(t-2\tau) \approx X_s(t)\cos(2\Omega_s\tau) - P_s(t)\sin(2\Omega_s\tau)$. As stated after equation~\eqref{fig_si_exp_setup_details}, the interference contrast of the spin signal when varying $\phi$ should not be diminished by optical loss because it affects the laser and quantum fields in the same way. Rather, optical delay $\tau$, imperfect optical polarization and differences in laser-atom mode-matching between the two passes are expected to reduce the contrast. From equation~\eqref{eq_x_output_field} it is calculated that the root-mean-square spin signal at the output is modulated by
\begin{equation}
\epsilon(\phi) = \sqrt{1 + \cos^2(\phi) + 2\cos(\phi)\cos(2\Omega_s\tau)},
\end{equation}
due to interference via the loop.

Fig.~\ref{fig_5}E shows the measured root-mean-squared spin signal in $X_L^\mathrm{(out)}$ for three different configurations. Two traces correspond to the double-pass atom-light interface with loop phases of $\phi = 0$ and $\phi = \pi$. The third trace shows the spin signal for a single pass interaction which is realized by moving the laser beam away from the atomic cloud in the second pass. The data clearly show a strong suppression of the spin signal for $\phi = \pi$ as compared to $\phi = 0$. Fitting the traces with an exponential decay including an initial detector rise time ($1/e$-time 10~$\mu$s) allows us to extract the amplitudes as well as the spin decay rates. First, we note that the double-pass signal for $\phi = 0$ is $3.3$ times larger than the single-pass output, which indicates a $1.6$-fold enhancement of the scattering efficiency in the presence of the second laser beam. Compared to $\phi = 0$, the spin signal at $\phi=\pi$ is suppressed by a factor $14$. This value is in good agreement with $\epsilon(0)/\epsilon(\pi) \approx 12$ for $2\Omega_s\tau = 0.17$. In this measurement, optical delay is only due to optical path length of about $4$~m because the cavity is off-resonant. 

Next, we discuss the effect of the loop phase on spin damping. The fitted (energy) damping rates are $\gamma_{s,\phi = 0} = 2\pi \times 3.1$~kHz in double pass with $\phi = 0$, $\gamma_{s,\phi = \pi} = 2\pi \times 1.9$~kHz in double pass with $\phi = \pi$, and $\gamma_{s,1} = 2\pi \times 0.45$~kHz in single pass. The broadening effect is also very clearly observed in the power spectra of the spin signals (Fig.~\ref{fig_5}F). The spin linewidths extracted from Lorentzian fits to the power spectra for single pass and double pass $\phi = \pi$ agree reasonably well with the damping rates. Only for double pass with $\phi = 0$, the decay is quite non-exponential such that the spectrum is not well fitted by a Lorentzian lineshape.

The increased damping rate in double pass is expected due to the enhanced spontaneous scattering rate at almost four times the optical intensity. Due to optical loss, the second beam has only about 80\% the optical power as the first beam, which should lead to a damping rate of $2\pi\times 1.5$~kHz. However, the measured damping rates in double pass are higher than that. This effect is likely explained by either additional broadening due to light-mediated self-interaction via the loop, or due to inhomogeneous optical light shifts arising from the crossed laser beams. 

Although we do not have an independent measurement of the back-action introduced by the coupling beam, observing constructive and destructive interference of the output spin signal clearly indicates leakage of spin information to the environment is suppressed. From the expression for the output field \eqref{eq_si_output_field} it is clear that full cancelation of the spin signal can also occur for nonzero optical loss, if both the laser field and the quantum field undergo the same optical loss as in our experiment. Observing high contrast interference of the spin signal with a contrast $> 10$ is important because it means that to a high degree both laser fields couple to the same spin wave. Based on the optical roundtrip transmission of $\eta^4 = 0.65$, we estimate a  spin back-action reduction to about $1-\eta^4 = 0.35$. This is already sufficient to reach the quantum-coherent coupling regime as shown in the subsequent section. 


\section{Coupled dynamics}
\label{sec_suppl_4}

\subsection{Heisenberg-Langevin equations}
\label{sec_hle}
Using input-output theory we derive a set of Heisenberg-Langevin equations for the cascaded spin-membrane system that is sketched in Fig.~\ref{fig_si_cascade_1} \cite{Karg2019}. For convenience, we neglect losses in this treatment. The equations of motion for the spin and membrane position and momentum operators read 
\begin{eqnarray}
\dot{X}_m &=& \phantom{-}\Omega_m P_m,\\
\dot{P}_m &=& - \Omega_m X_m -\gamma_m P_m  - 2g X_s(t-\tau) - \sqrt{4\Gamma_m} X_L^\mathrm{(in)}(\zeta_2) - \sqrt{2\gamma_m} F_m ^{\mathrm{(th)}},\\
\dot{X}_s &=& \phantom{-}\Omega_s P_s,\\
\dot{P}_s &=& - \Omega_s X_s -\gamma_s P_s + 4\Gamma_s \sin(\phi) X_s(t-2\tau) + 2g \cos(\phi) X_m(t-\tau)\\
	& & \qquad - \sqrt{4\Gamma_s}\left[P_L^\mathrm{(in)}(\zeta_1) + \cos(\phi) P_L^\mathrm{(in)}(\zeta_3) - \sin(\phi) X_L^\mathrm{(in)}(\zeta_3)\right] - \sqrt{2\gamma_s} F_s^\mathrm{(th)}.\qquad \label{eq_si_spin_langevin_noise}
\end{eqnarray}
Here, $g = 2 \sqrt{\Gamma_m \Gamma_s}$ is the spin-membrane coupling strength, $\tau$ is the optical propagation delay between the systems which we assume to be equal for either direction, and $F_m^\mathrm{(th)}$ and $F_s^\mathrm{(th)}$ are mechanical and spin thermal noise terms, respectively. Each oscillator is also driven by optical vacuum noise of the input field quadratures $X_L^\mathrm{(in)}(\zeta_i), P_L^\mathrm{(in)}(\zeta_i)$ at the different locations $\zeta_i$ along the optical path. This leads to quantum back-action of the light that mediates the spin-membrane interaction onto the coupled systems. For the spin oscillator, the optical input terms at the two locations $\zeta_1$ and $\zeta_3$ interfere as can be seen directly in line \eqref{eq_si_spin_langevin_noise}. For the membrane there is no such interference as it interacts with the light field only once. Moreover, the two spin-light interactions also enable delayed light-mediated self-interaction of the spin. The effect of this is a modified frequency and linewidth since $X_s(t-2\tau) \approx X_s \cos(2\Omega_s\tau) - P_s\sin(2\Omega_s\tau)$. We thus have a spin frequency shift $\delta\Omega_s = 2\Gamma_s\sin(\phi)\cos(2\Omega_s\tau)$ and a shift of the damping rate $\delta\gamma_s = 4\Gamma_s\sin(\phi)\sin(2\Omega_s\tau)$. Since the atom-light coupling strength is inhomogeneous across the atomic ensemble, this can also lead to inhomogeneous broadening of the spin oscillator if $\phi \mod \pi \neq 0$.

In the following treatment, we assume $\phi$ to take only the discrete values $0,\pi$. Making a Fourier transform yields
\begin{eqnarray}
\chi_{m,0}(\omega)^{-1}X_m(\omega) + 2g e^{i\omega\tau} X_s(\omega) &=& - \sqrt{2\gamma_m} F_m^\mathrm{(tot)}(\omega)\\
\chi_{s,0}(\omega)^{-1}X_s(\omega) - 2g \cos(\phi) e^{i\omega\tau} X_m(\omega) &=& -\sqrt{2\gamma_s} F_s^\mathrm{(tot)}(\omega),
\end{eqnarray}
with bare (uncoupled) susceptibility defined as
\begin{equation}
\chi_{i,0}(\omega) = \frac{\Omega_i}{\Omega_i^2 -\omega^2 - i \omega \gamma_i}.
\end{equation}
and the combined thermal and optical force terms
\begin{eqnarray}
F_m^\mathrm{(tot)}(\omega) &=& F_m^\mathrm{(th)}(\omega) + \sqrt{\frac{2\Gamma_m}{\gamma_m}} e^{i\omega\tau} X_L^\mathrm{(in)}(\omega)\\
F_s^\mathrm{(tot)}(\omega) &=& F_s^\mathrm{(th)}(\omega) + \sqrt{\frac{2\Gamma_s}{\gamma_s}} \left[1 + \cos(\phi)e^{i2\omega\tau}\right]P_L^\mathrm{(in)}(\omega)
\end{eqnarray}

Solving for $X_m$, $X_s$ yields the solutions
\begin{eqnarray}
X_m(\omega) &=& \chi_{m,\mathrm{eff}}(\omega)\left[-\sqrt{2\gamma_m} F_m^\mathrm{(tot)}(\omega) + 2g e^{i\omega\tau}\sqrt{2\gamma_s} \chi_{s,0}(\omega) F_s^\mathrm{(tot)}(\omega)\right],\\
X_s(\omega) &=& \chi_{s,\mathrm{eff}}(\omega) \left[-\sqrt{2\gamma_s} F_s^\mathrm{(tot)}(\omega) - 2g \cos(\phi) e^{i\omega\tau}\sqrt{2\gamma_m} \chi_{m,0}(\omega) F_m^\mathrm{(tot)}(\omega)\right],
\end{eqnarray}
where we have used the effective susceptibilities of the membrane
\begin{eqnarray}
\chi_{m,\mathrm{eff}}(\omega)^{-1} &=& \chi_{m,0}(\omega)^{-1} + 4g^2 \cos(\phi) e^{i2\omega\tau} \chi_{s,0}(\omega),\\
\chi_{s,\mathrm{eff}}(\omega)^{-1} &=& \chi_{s,0}(\omega)^{-1} + 4g^2 \cos(\phi) e^{i2\omega\tau} \chi_{m,0}(\omega).
\end{eqnarray}

For the fits of the normal mode splittings in Fig.~\ref{fig_2}A and C of the main text we use the fitting function $a|\chi_{m,\mathrm{eff}}(\omega)|$ with scaling factor $a$ and other fit parameters being $g, \gamma_m, \gamma_s, \Omega_s, \tau$. The argument of $\chi_{m,\mathrm{eff}}$ gives the phase response which is plotted in Figs.~\ref{fig_2}B and D together with the experimental data. The fit of the normal-mode splitting with our theoretical model yields a delay of $\tau=15$~ns. This value is close to a calculated value of $2/\kappa + d/c = 12$~ns based on the cavity linewidth $\kappa = 2\pi\times 63$~MHz (full width at half-maximum) and optical propagation distance $d = 2$~m, $c$ being the speed of light.

\subsection{Normal modes}

\begin{figure}[tb!]
\centering
\includegraphics[width = 0.66\linewidth]{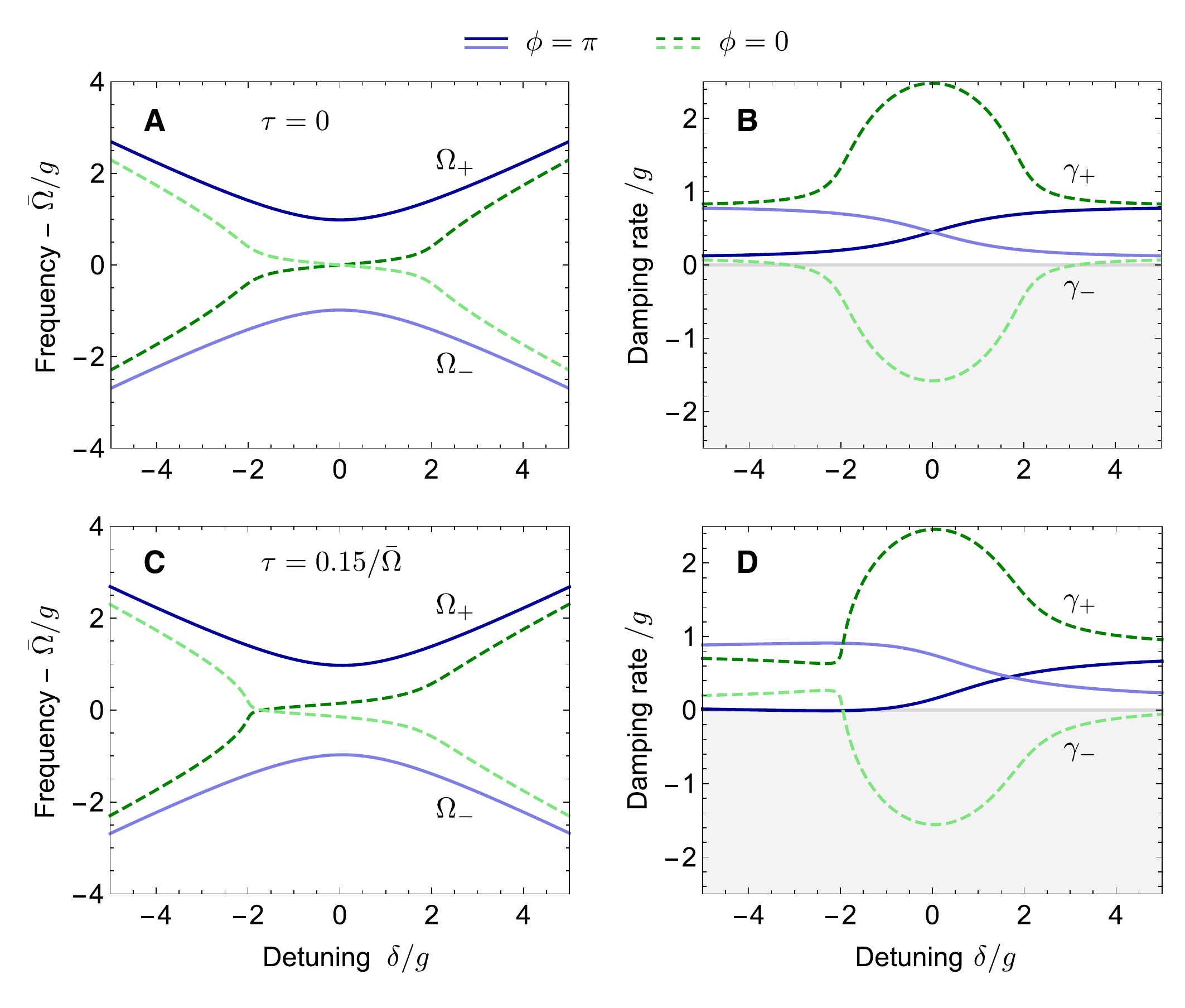}
\caption{Normal mode frequencies (A,C) and damping rates (B,D) as a function of oscillator detuning. The solid blue lines correspond to the Hamiltonian coupling with $\phi = \pi$ while the dashed green lines are for $\phi = 0$. The delay is chosen to be $\tau = 0$ in A,B and $\tau = 0.15/\bar{\Omega}$ in C,D.}
\label{fig_si_normal_mode_eigenvalues}
\end{figure}

The normal mode frequencies and damping rates can be obtained more easily from an analysis using the rotating wave approximation, which is a very good approximation because $g,\gamma_i < \Omega_i/10^2$ in the experiment. We perform here the calculation for the positive spin oscillator $\Omega_s > 0$. The coupled equations of motion for the mode operators $b_m$ and $b_s$ in a rotating frame at the center frequency $\bar{\Omega} = (\Omega_m + \Omega_s)/2$ read
\begin{eqnarray}
\dot{b}_m &=& \left(+ i \frac{\delta}{2} - \frac{\gamma_m}{2}\right) b_m - i g e^{i\Omega_s \tau} b_s - i \sqrt{\gamma_m} F_m,\\
\dot{b}_s &=& \left(- i \frac{\delta}{2} - \frac{\gamma_s}{2}\right) b_s + i g \cos(\phi) e^{i\Omega_m \tau} b_m - i \sqrt{\gamma_s} F_s,
\end{eqnarray}
where we defined the spin-membrane detuning $\delta = \Omega_s - \Omega_m$. For the inverted spin oscillator one would replace $b_s \to b_s^\dag$, $\Omega_s \to - \Omega_s$ and in the second line $g\to-g$. Solving for the eigenvalues of the dynamical matrix gives the frequencies $\Omega_{\pm}$ and damping rates $\gamma_{\pm}$ via the relation
\begin{equation}
\Omega_{\pm} + i\frac{\gamma_{\pm}}{2} = \bar{\Omega} + i\frac{\gamma_m + \gamma_s}{4} \pm \sqrt{\left(\frac{\delta}{2} + i \frac{\gamma_m - \gamma_s}{4}\right)^2 - g^2 e^{i2\bar{\Omega}\tau} \cos(\phi)}.
\end{equation}
For illustration, the normal mode frequencies and damping rates are plotted in Figs.~\ref{fig_si_normal_mode_eigenvalues}A,C and \ref{fig_si_normal_mode_eigenvalues}B,D, respectively, for a choice of parameters $\gamma_m / g = 0.1$, $\gamma_s / g = 1$ that reflect the situation in the experiment. In Figs.~\ref{fig_si_normal_mode_eigenvalues}A,B the delay is set to $\tau = 0$ while in Figs.~\ref{fig_si_normal_mode_eigenvalues}C,D we choose $\bar{\Omega}\tau = 0.15$ as in the experiment. 

We first discuss the Hamiltonian coupling case ($\phi = \pi$). In a standard instantaneous coupling situation ($\tau = 0$), the normal modes exhibit an avoided crossing at $\delta = 0$ with a splitting of $\Omega_+ - \Omega_- = 2\sqrt{g^2 - (\gamma_m - \gamma_s)^2/16} \approx 2 g$.  The damping rates are then exactly $\gamma_\pm = (\gamma_m + \gamma_s)/2$ the average of the individual damping rates. With increasing detuning, the normal mode frequencies and damping rates smoothly transform into those of the uncoupled modes (see Fig.~\ref{fig_si_normal_mode_eigenvalues}B).

With a small delay $\bar{\Omega}\tau = 0.15$, the situation changes. While the normal mode frequencies are hardly affected, the damping rates change significantly. Looking at Fig.~\ref{fig_si_normal_mode_eigenvalues}D (solid lines), we see that the delay induces an asymmetry of the damping rates. The point where both damping rates are equal shifts to positive detunings $\delta > 0$. Moreover, at detunings $\delta < 0$ we see an increased splitting between the damping rates leading to one of them closely approaching zero and even becoming slightly negative for certain detunings. This leads to an instability of the coupled dynamics as observed in Fig.~\ref{fig_5}A of the main text.

For dissipative coupling ($\phi = 0$) the normal modes exhibit level attraction (see Fig.~\ref{fig_si_normal_mode_eigenvalues}A, dashed lines). The modes become degenerate only at a single point because of a difference in their individual damping rates. Otherwise the normal modes would merge in the full range $|\delta|<2g$. The two normal modes exhibit a splitting in terms of their damping rates (see Fig.~\ref{fig_si_normal_mode_eigenvalues}B) leading to one strongly damped mode with $\gamma_+ \approx (\gamma_m + \gamma_s)/2 + 2 g > 0$ and one amplified mode with $\gamma_- \approx (\gamma_m + \gamma_s)/2 - 2 g< 0$. In the parametric-gain configuration ($\phi = \pi$, $\Omega_s = -\Omega_m$) these modes correspond to the squeezed and anti-squeezed modes, respectively. Like in the Hamiltonian coupling, a finite delay introduces an asymmetry both of the degeneracy point of the mode frequencies as well as in the detuned damping rates (see Figs.~\ref{fig_si_normal_mode_eigenvalues}C,D).

In Fig.~\ref{fig_5} of the main text we globally fit a function $a + b|\chi_{m,\mathrm{eff}}(\omega)|^2$ to the experimental data in panels A and D that exhibit avoided crossings. The fits yields the coupling strength $g$, detuning $\delta$, linewidths $\gamma_{s/m}$ and delay $\tau$ which we use to calculate the theoretical normal mode frequencies $\Omega_\pm$ which are drawn as dashed lines. The data in panels B and C are not fitted, because the dynamics are unstable in these configurations and thus do not reach a steady state. Instead, to calculate the theory curves for B and C we use the same parameters obtained for the fits to the data in A and D, respectively.

\subsection{Simulation of the covariance dynamics}

To simulate the two-mode thermal noise squeezing in figure~4B, we solve the time-dependent Lyapunov equation \cite{Hofer2015,Karg2019}
\begin{equation}
\dot{\Sigma} = F \Sigma + \Sigma F^T + N,
\end{equation}
for the symmetrized covariance matrix $\Sigma_{jk} = \langle Q_j Q_k + Q_k Q_j\rangle/2 - \langle Q_j\rangle \langle Q_k \rangle$, where $Q = (X_m, P_m, X_s, P_s)$. The drift matrix $F$ and diffusion matrix $N$ are obtained from the master equation \eqref{eq_master_eq_spinmembrane_loss} written in the form
\begin{equation}
\dot{\rho} = - \sum_{j, k} A_{jk} [Q_j, Q_k \rho] + \mathrm{h.c.},
\end{equation}
The expressions for $F$ and $N$ are \cite{Hofer2015,Karg2019}
\begin{equation}
F = 2 J \im\left\{A\right\},\quad N = J \re\left\{A + A^T\right\} J^T,
\end{equation}
with commutator matrix $J_{jk} = -i[Q_j, Q_k]$. In the simulation, we choose the experimental parameters as listed in table~\ref{tab_parameters_cov}. They correspond to $2g = 2\pi \times 5.2$~kHz due to a slight reduction of the spin pumping efficiency in the inverted configuration. Moreover, we find best agreement with $\gamma_s = 2\pi\times 1$ kHz, implying that the spin linewidth observed in the spectroscopy is mostly due to inhomogeneous broadening. To simulate detector noise, we add $\bar{n}_\mathrm{det} = 6\times 10^3$ to the covariance entries $\langle X_s^2\rangle - \langle X_s\rangle^2$ and $\langle P_s^2\rangle - \langle P_s\rangle^2$.

\begin{table}[tb!]
\caption{Experimental parameters used for the simulation of two-mode thermal noise squeezing.}
\centering
\begin{tabular}{c | r}
Parameter	& Value\\\hline\hline
$\Omega_m$	& $2\pi \times 1.957$ MHz\\
$\Gamma_m$	& $2\pi \times 7.5$ kHz\\
$\gamma_m$	& $2\pi \times 0.4$ kHz\\
$\bar{n}_m$	& $1.5\times 10^4$\\\hline
$\Omega_s$	& $2\pi \times 1.957$ MHz\\
$\Gamma_s$	& $2\pi \times 0.43$ kHz\\
$\gamma_s$	& $2\pi \times 1$ kHz\\
$\bar{n}_s$	& $0$\\\hline
$\eta$		& $0.9$\\
$\phi$		& $\pi$\\
\end{tabular}
\label{tab_parameters_cov}
\end{table}

\subsection{Reaching the quantum regime}

In this section we estimate the performance of the presented experimental setup in the quantum regime. Our criterion for quantum coherent coupling in the Hamiltonian coupling ($\phi = \pi$) is the ability to achieve entanglement using the parametric-gain interaction.

Next to the coherent coupling, the effective master equation for the light-coupled system
\begin{eqnarray}
\dot{\rho} &=& \frac{1}{i\hbar} [H_0 + H_\mathrm{eff},\rho]\\
	& & + \sum_{i=s,m}\left(\gamma_{i,0}(\bar{n}_i + 1)\mathcal{D}[b_i]\rho + \gamma_{i,0}\mathcal{D}[b_i^\dag]\rho \right)\\
	& & + \sum_{i=s,m}\left(\gamma_{i,\mathrm{ba}}\mathcal{D}[b_i]\rho + \gamma_{i,\mathrm{ba}}\mathcal{D}[b_i^\dag]\rho \right)
\end{eqnarray}
features various dissipative terms. Apparently, for each system there are thermal (second line) and an optical (third line) decoherence processes. The thermal decoherence rates are given by $\gamma_{i,\mathrm{th}} = \gamma_{i,0} (\bar{n}_i + 1/2)$, where $\gamma_{i,0}$ is the damping rate and $\bar{n}_i$ is the thermal bath occupation number of system $i$. The optical back-action decoherence rates are $\gamma_{m,\mathrm{ba}} = \eta^2 \Gamma_m$ for the membrane and $\gamma_{s,\mathrm{ba}} = (1- \eta^4) \Gamma_s$ for the spin, where we have assumed an average amplitude transmission coefficient $\eta$ per path. In the following, we define the total decoherence rates $\gamma_{i,\mathrm{tot}}$ for each oscillator as the sum of their independent thermal and optical back-action decoherence rates, i.e. $\gamma_{i,\mathrm{tot}} = \gamma_{i,\mathrm{th}} + \gamma_{i,\mathrm{ba}}$. 

For Gaussian states we can quantify entanglement as a violation of the non-separability criterion \cite{Duan2000a,Simon2000}
\begin{equation}
\xi := \langle X_-^2\rangle + \langle P_+^2\rangle < 1,
\label{eq_si_entanglement}
\end{equation}
where we defined the collective quadratures $X_\pm = (X_s \pm X_m)/\sqrt{2}$, $P_\pm = (P_s \pm P_m)/\sqrt{2}$. 

We now calculate the amount of entanglement realized by the two-mode suqeezing interaction in the configuration $\phi = \pi$, $\Omega_s = -\Omega_m$. Here, we choose a phase convention such that $H_\mathrm{PG} = i\hbar g (b_s^\dag b_m^\dag - b_s b_m)$. Applying a rotating-wave-approximation and transforming into the basis $X_\pm$, $P_\pm$, we derive a set of coupled differential equations for the entries of the spin-membrane covariance matrix, i.e.
\begin{eqnarray}
\frac{d}{dt} \langle X_+^2\rangle &=& + \frac{4 g - \gamma_{s,0} - \gamma_{m,0}}{2} \langle X_+^2\rangle + \frac{\gamma_{s,0} - \gamma_{m,0}}{2} \langle X_{+} X_{-}\rangle + \frac{\gamma_{s,\mathrm{tot}} + \gamma_{m,\mathrm{tot}}}{2}\\
\frac{d}{dt} \langle X_-^2\rangle &=& - \frac{4 g + \gamma_{s,0} + \gamma_{m,0}}{2} \langle X_-^2\rangle + \frac{\gamma_{s,0} - \gamma_{m,0}}{2} \langle X_{+} X_{-}\rangle + \frac{\gamma_{s,\mathrm{tot}} + \gamma_{m,\mathrm{tot}}}{2}\\
\frac{d}{dt} \langle X_+ X_-\rangle &=& - \frac{\gamma_{s,0} + \gamma_{m,0}}{2} \langle X_+ X_-\rangle + \frac{\gamma_{s,0} - \gamma_{m,0}}{4} (\langle X_{+}^2\rangle + \langle X_{-}^2\rangle) - \frac{\gamma_{s,\mathrm{tot}} - \gamma_{m,\mathrm{tot}}}{2}\qquad
\end{eqnarray}
Note that in rotating-wave approximation, $\langle P_\pm^2\rangle = \langle X_\mp^2\rangle$. The above equations imply that $X_{-}$ and $P_+$ are squeezed while $X_+$ and $P_-$ are anti-squeezed. If the damping rates $\gamma_{i,0}$ or total decoherence rates $\gamma_{i,\mathrm{tot}}$ are unequal, the squeezed and anti-squeezed quadratures deviate slightly from $X_-,P_+$ and $X_+,P_-$, respectively, as we see from the terms involving the covariance $\langle X_+ X_-\rangle$. Thermal and optical back-action noise appears in form of the constant terms $\gamma_{s,\mathrm{tot}} + \gamma_{m,\mathrm{tot}}/2$.
Assuming $(\gamma_{s,0} - \gamma_{m,0})/ 4 g \ll 1$ we find that in steady state,
\begin{equation}
\langle X_-^2\rangle \approx \frac{\gamma_{s,\mathrm{tot}} + \gamma_{m,\mathrm{tot}}}{4 g + \gamma_{s,0} + \gamma_{m,0}} + \mathcal{O}\left(\frac{\gamma_{s,0} - \gamma_{m,0}}{4g}\right)
\end{equation}
Entanglement in terms of equation~\eqref{eq_si_entanglement} is equivalent to reduction of $\langle X_{-}^2\rangle$ below $1/2$. Consequently, to generate entanglement the coupling strength $g$ needs to exceed the average of all decoherence rates on both the mechanical and spin system, i.e. $2g  > \gamma_{s,\mathrm{tot}} + \gamma_{m,\mathrm{tot}}$, as
\begin{equation}
\xi = \left(\frac{1}{1 + 2\bar{n}_\mathrm{eff}} + C\right)^{-1}
\end{equation}
where $C = 2 g / (\gamma_{m,\mathrm{tot}} + \gamma_{s,\mathrm{tot}})$ is a quantum cooperativity parameter and $\bar{n}_\mathrm{eff} = (\gamma_{m,\mathrm{tot}} + \gamma_{s,\mathrm{tot}})/(\gamma_{m,0} + \gamma_{s,0}) - 1/2$ is the average occupation number of the collective mode. The entanglement criterion $\xi < 1$ thus requires $C>1$, which is equivalent to the condition for quantum coherent coupling of ref.~\cite{Verhagen2012}.

With a meaningful criterion for quantum coherent coupling, we now estimate the required system parameters to reach this regime. Clearly, thermal noise is the largest contribution to mechanical decoherence. For the current room temperature ($T_m = 295$~K) implementation with $\bar{n}_m \approx  k_B T_m / (\hbar \Omega_m )= 3 \times 10^6$ and a mechanical quality factor of $Q_m = 1.3\times 10^6$, we have $\gamma_{m,\mathrm{th}} \approx \gamma_{m,0} \bar{n}_m \approx 2\pi \times 6$~MHz. Lowering the bath temperature to $T_m = 5$~K by cooling with liquid helium and increasing $Q_m$ to $5\times 10^7$ could reduce the thermal decoherence rate to $\gamma_{m,\mathrm{th}} \approx 2\pi \times 2$~kHz. Such quality factors have recently been demonstrated by soft-clamping of mechanical modes in a high-stress silicon nitride membrane \cite{Tsaturyan2017}. At this level, $\gamma_{m,\mathrm{th}}$ would be of similar magnitude or even lower than the optomechanical measurement rate $\Gamma_m \approx 2\pi\times 8$~kHz in the current experiment, which is also equal to the optical back-action rate $\gamma_{m,\mathrm{ba}}$ for the membrane. Hence, the optomechanical system would reach the regime of large quantum cooperativity $\Gamma_m/\gamma_{m,\mathrm{th}} > 1$ where mechanical fluctuations are dominated by optical back-action instead of thermal noise.

Tuning of the atom-light interaction is achieved by controlling the laser detuning $\Delta_a$ from the atomic transition. Both the spin measurement rate $\Gamma_s$ and the spontaneous photon scattering rate $\gamma_\mathrm{sc}$ scale with $\Phi_L\Delta_a^{-2}$. Consequently, $g\propto\sqrt{\Gamma_s} \propto \sqrt{\Phi_L}\Delta_a^{-1}$ can be increased relative to $\gamma_\mathrm{sc}$ at large detuning. Since the laser input power $\Phi_L$  also affects the optomechanical measurement rate it is kept fixed in this optimization. For a highly spin-polarized cold atomic ensemble, one can assume $\bar{n}_s \approx 0$ which eliminates thermal noise. The spin back-action rate $\gamma_{s,\mathrm{ba}} = (1-\eta^4)\Gamma_s$ is suppressed due to destructive interference in the loop. In the experiment, a system-to-system optical power transmission of $\eta^2 \approx 0.8$ is achieved, resulting in optical back-action suppression down to $1-\eta^4 \approx 0.35$.

For quantum coherent spin-membrane coupling we need to make $g = (\eta^2 +\eta^4)\sqrt{\Gamma_s \Gamma_m}$ larger than $\gamma_{m,\mathrm{tot}}$ and $\gamma_{s,\mathrm{tot}}$. Since there is no back-action cancellation for the membrane, the requirement $g > \gamma_{m,\mathrm{tot}}$ leads to $\Gamma_s \geq \Gamma_m$. This constraint limits the maximum possible laser-atom detuning and therefore entails a minimum spontaneous scattering rate, which reduces spin coherence. A large spin cooperativity is thus crucial for strong coupling in the hybrid system.

\begin{figure}[tb!]
\centering
\includegraphics[width = 0.66\linewidth]{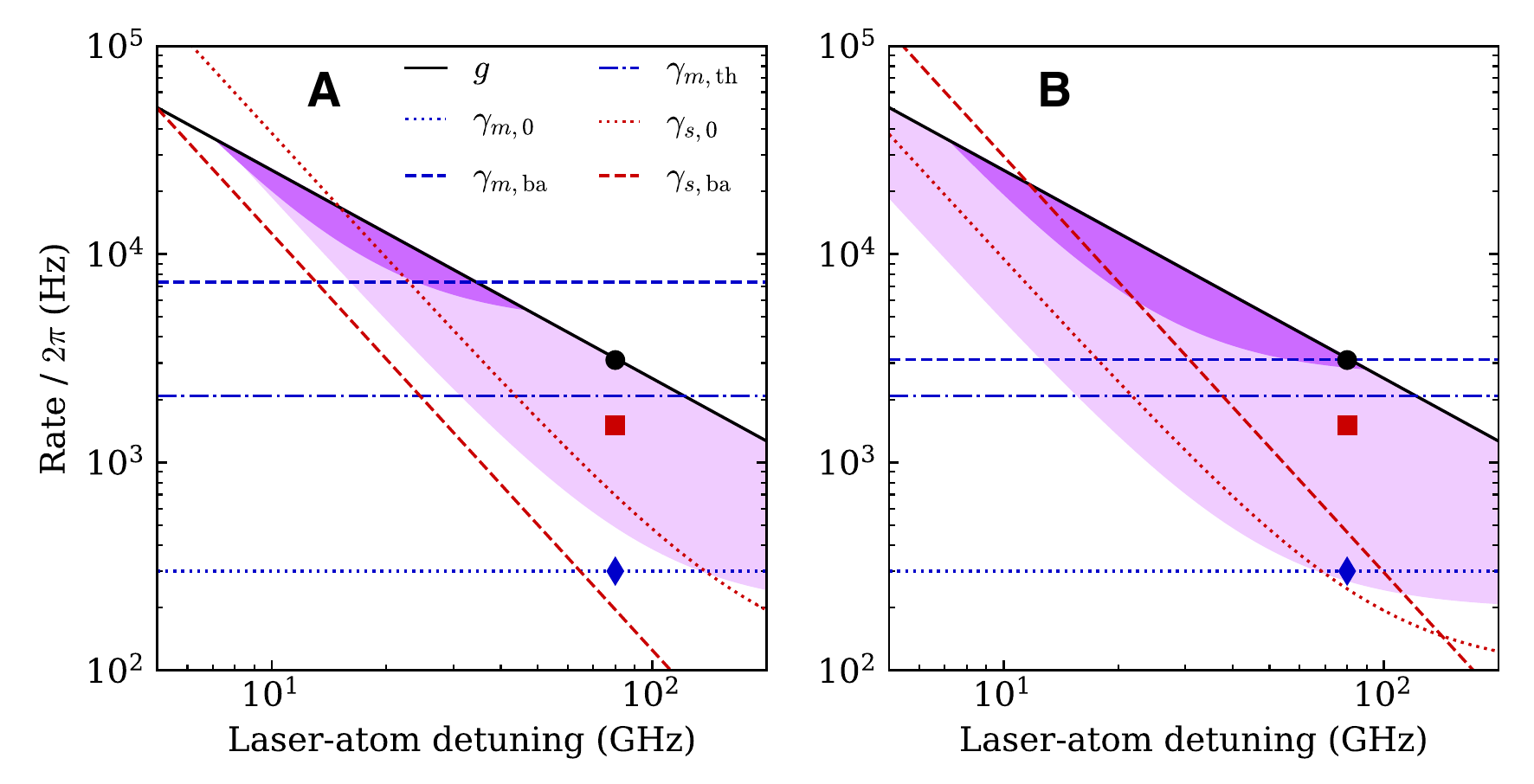}
\caption{Rates of the coupled spin-membrane system as a function of the laser-atom detuning $\Delta_a$. (A), Configuration with quantum noise cancelation on spin. (B), Configuration with quantum noise cancelation on membrane. The lines are calculated rates based on independently measured experimental parameters. The points correspond to experimentally observed parameters for $g$ (black circle), $\gamma_{s,0}$ (red square) and $\gamma_{m,0}$ (blue diamond). The light purple shaded area corresponds to the margin between $g$ and the mean damping rate $(\gamma_{s,0} + \gamma_{m,0})/2$, indicating strong coupling. The margin between $g$ and the total decoherence rate $(\gamma_{s,\mathrm{tot}} + \gamma_{m,\mathrm{tot}})/2$ for quantum coherent coupling is given by the dark purple area.}
\label{fig_si_rates}
\end{figure}

In Fig.~\ref{fig_si_rates} we show calculated rates of the spin-membrane system as a function of the laser-atom detuning $\Delta_a$. Here, we assume $Q_m = 5\times 10^7$ and a mechanical bath temperature of $T_m = 5$~K. Together with modest optomechanical damping such that $\gamma_{m,0} = 2\pi \times 300$~Hz, this would result in an effective phonon occupation of $\bar{n}_m \approx 7$. 

Fig.~\ref{fig_si_rates}A shows the calculated rates for an experimental setup with a loop on the spin, suppressing its optical back-action noise. The black circle is the experimentally determined value of $g = 2\pi\times 3.1$~kHz, obtained at $\Delta_{a}/2\pi = -80~$GHz, which agrees very well with the calculated curve. The mechanical damping rate $\gamma_{m,0} = 2\pi\times300$~Hz is indicated by the blue diamond. The experimentally determined spin damping rate (red square) $\gamma_{s,0} = 2\pi \times 1.5$~kHz is a factor of two larger than the theoretical value $2\pi\times 0.7$~kHz which means that other decoherence effects are present. We note that the spin linewidth ($4~$kHz) extracted from the normal-mode splitting data is even larger than this value. A possible explanation for this is very likely the more complicated atom-light interface with two crossed laser beams. This could lead to light-induced broadening caused by polarization gradients or atomic self-interaction. 

The margin between $g$ and the average damping rate $(\gamma_{s,0} + \gamma_{m,0})/2$ for strong coherent coupling is colored light purple. Dark purple denotes the margin between $g$ and the total decoherence rate $(\gamma_{s,\mathrm{tot}} + \gamma_{m,\mathrm{tot}})/2$ for quantum coherent coupling. Obviously, strong coupling is easier to achieve than quantum coherent coupling, for which there is only a narrow parameter window. Clearly, both the spin damping rate and the mechanical back-action rate are strongly limiting the achievable cooperativity $C$. Hence, we also show the calculated rates in the cascaded coupling scenario with a loop on the optomechanical system instead of the atomic ensemble (see Fig.~\ref{fig_si_rates}B). Here, optical back-action is suppressed on the membrane, but not on the spin. Moreover, the spin's damping rate decreases because of the reduced scattering rate in a single laser beam. For the membrane, increase of the damping rate can be compensated by a smaller laser-cavity detuning. In this situation the cooperativity $C$ scales more favorably with detuning, leading to a wider region where quantum coherent coupling is possible. Finally, we remark that this situation would improve even more if quantum back-action was canceled on both systems. This requires double pass light-matter interactions on both systems \cite{Karg2019} and is therefore more difficult to implement, but possible in principle.

\end{document}